\pdfoutput=1

\documentclass[11pt,twoside,a4paper,cmspaper,final,collab]{cms-tdr}
\def\svnVersion{188638}\def\cmsCernNo{2012-343}\def\cmsCernDate{\today}\def\cmsMessage{Submitted to Physical Review D}

\begin{document}\cmsNoteHeader{QCD-11-004}

\hyphenation{had-ron-i-za-tion}
\hyphenation{cal-or-i-me-ter}
\hyphenation{de-vices}
\RCS$HeadURL: svn+ssh://svn.cern.ch/reps/tdr2/papers/QCD-11-004/trunk/QCD-11-004.tex $
\RCS$Id: QCD-11-004.tex 183674 2013-04-30 04:19:15Z gmavroma $
\newlength\cmsFigWidth
\ifthenelse{\boolean{cms@external}}{\setlength\cmsFigWidth{0.85\columnwidth}}{\setlength\cmsFigWidth{0.4\textwidth}}
\ifthenelse{\boolean{cms@external}}{\providecommand{\cmsLeft}{top}}{\providecommand{\cmsLeft}{left}}
\ifthenelse{\boolean{cms@external}}{\providecommand{\cmsRight}{bottom}}{\providecommand{\cmsRight}{right}}
\ifthenelse{\boolean{cms@external}}{\providecommand{\cmsLLeft}{Top}}{\providecommand{\cmsLLeft}{Left}}
\ifthenelse{\boolean{cms@external}}{\providecommand{\cmsRRight}{Bottom}}{\providecommand{\cmsRRight}{Right}}
\newcommand{\Mjj}{\ensuremath{M_\text{jj}}\xspace}
\providecommand{\suppMaterial}{the supplemental material [URL will be inserted by publisher]}
\cmsNoteHeader{QCD-11-004}%
\title{Measurements of differential jet cross sections in proton-proton collisions at
  \texorpdfstring{$\sqrt{s}=7\TeV$}{sqrt(s) = 7 TeV} with the CMS detector}

\author{The CMS Collaboration}

\date{\today}

\abstract{Measurements of inclusive jet and dijet production cross
  sections are presented. Data from LHC proton-proton collisions at
  $\sqrt{s}=7\TeV$, corresponding to 5.0\fbinv of integrated
  luminosity, have been collected with the CMS detector. Jets are
  reconstructed up to rapidity 2.5, transverse momentum 2\TeV, and
  dijet invariant mass 5\TeV, using the anti-\kt clustering
  algorithm with distance parameter $R=0.7$. The measured cross
  sections are corrected for detector effects and compared to
  perturbative QCD predictions at next-to-leading order, using five
  sets of parton distribution functions.}

\hypersetup{%
pdfauthor={CMS Collaboration},
pdftitle={Measurements of differential jet cross sections in proton-proton collisions at
  sqrt(s)=7 TeV with the CMS detector},
pdfsubject={CMS},
pdfkeywords={CMS, physics, QCD, jets}}

\maketitle

\section{Introduction}

Events with high transverse momentum jets in proton-proton collisions
are described by quantum chromodynamics (QCD) in terms of
parton-parton scattering, where the outgoing scattered partons
manifest themselves as hadronic jets. Measurements of the inclusive jet and dijet
cross sections can be used to test the predictions of
perturbative QCD, constrain parton distribution functions (PDFs) of
the proton, differentiate among PDF sets, and look for possible
deviations from the standard model.

In this paper, measurements of the double-differential inclusive
jet ($\Pp + \Pp \to \text{jet} + X$) and dijet ($\Pp + \Pp \to \text{jet} +
\text{jet} + X$) production cross sections are reported as functions of jet
rapidity $y$ and either jet transverse momentum \pt or dijet
invariant mass $\Mjj$, at $\sqrt{s}=7\TeV$. The data were
collected with the Compact Muon Solenoid (CMS) detector at the CERN
Large Hadron Collider (LHC) during the 2011 run and correspond to an
integrated luminosity of 5.0\fbinv, two orders of magnitude larger
than the published LHC results from the 2010
run~\cite{ATLAS,CMSDijets,CMSInclusive}. Jets are reconstructed up to rapidity 2.5, transverse momentum
2\TeV, and  dijet invariant mass 5\TeV.
The measured cross sections
are corrected for detector effects and compared to the next-to-leading-order QCD predictions.
\section{Apparatus}

The CMS coordinate system has its origin at the center of the
detector, with the $z$ axis pointing along the direction of the
counterclockwise beam. The azimuthal angle is denoted as $\phi$, the
polar angle as $\theta$, and the pseudorapidity is defined as
$\eta=-\ln\left[\tan\left(\theta/2\right)\right]$. The central feature
of the CMS detector is a superconducting solenoid, of 6\,m internal
diameter, that produces an axial magnetic field of 3.8\,T.  Within the
field volume are a silicon pixel and strip tracker, a lead tungstate
crystal electromagnetic calorimeter (ECAL) and a brass/plastic
scintillator hadronic calorimeter. Outside the field volume and in the
forward region ($3 < |\eta| < 5$) is an iron/quartz-fiber
hadronic calorimeter. Muons are measured in gas ionization detectors
embedded in the steel return yoke outside the solenoid, in the
pseudorapidity range $|\eta| < 2.4$. A detailed description of the CMS
apparatus can be found in Ref.~\cite{CMS}.
\section{Jet reconstruction}

The rapidity $y$ and the transverse momentum \pt of a jet with energy
$E$ and momentum $\vec{p}=(p_x,p_y,p_z)$ are defined as
$y=(1/2)\ln\left[(E+p_z)/(E-p_z)\right]$ and $\pt=\sqrt{p_x^2+p_y^2}$.
The inputs to the jet clustering algorithm are the
four-momentum vectors of the reconstructed particle candidates. Each
such candidate is constructed using the particle-flow
technique~\cite{PFLOW}, which combines the information from several
subdetectors and is calibrated to account for the nonlinear and
nonuniform response of the CMS calorimetric system to hadrons.
Jets are reconstructed using the anti-\kt clustering algorithm~\cite{AKT}
with distance parameter $R=0.7$. The clustering is performed using
four-momentum summation with the {\sc Fastjet}
package~\cite{FASTJET}, where the chosen distance parameter allows for
the capture of most of the parton shower and improves the dijet mass
resolution with respect to smaller sizes. The total transverse energy
$\Sigma\,\ET$ and missing transverse energy $\VEtmiss$ are
used in the event selection and are derived from the reconstructed
particle-flow objects. They are defined as $\Sigma\,\ET =
\sum_{i}{{\et}_i}$, with $E_{\text{T}i} = E_i \sin\theta_i$, and
$\VEtmiss=-\sum_{i}{\left(E_i\sin\theta_i\cos\phi_i\hat{\mathbf{x}} +
    E_i\sin\theta_i\sin\phi_i\hat{\mathbf{y}}\right)}$, where the sum
refers to all particle candidates and
$\hat{\mathbf{x}},\hat{\mathbf{y}}$ are the unit vectors in the
direction of the $x$ and $y$ axes.

The reconstructed jets require a small additional energy correction,
mostly due to thresholds on reconstructed tracks and clusters in the
particle-flow algorithm, and various reconstruction
inefficiencies. These jet energy corrections are derived using simulated
events, generated by \PYTHIA{6} (version 6.4.22)~\cite{PYTHIA} and
processed through the CMS detector simulation based on \GEANTfour~\cite{GEANT4}, and in situ measurements with dijet,
photon+jet, and Z+jet events~\cite{JES}. These jet energy corrections
correct reconstructed jets to the hadron level, as opposed to the
parton level. An offset correction is also applied to
account for the extra energy from additional
proton-proton interactions within the same or neighboring bunch
crossings (in-time and out-of-time pileup)~\cite{JES}. The pileup
effects are important for the lowest-\pt jets ($10\%$ jet energy scale
correction and 1\% systematic uncertainty~\cite{JES2011} for jets
with $\pt\sim 100\GeV$) and progressively decrease with jet \pt. For
jets with $\pt>200\GeV$ the pileup effects are negligible. The jet
energy correction depends on $\eta$ and \pt of the jet, and is applied
as a multiplicative factor to the jet four-momentum vector. The factor
is typically between 1.0 and 1.2 and is approximately uniform in $\eta$.
For a jet $\pt=100\GeV$ the factor is 1.1, decreasing towards 1.0 with
increasing \pt. The typical jet \pt resolution is $10\%$ at
$\pt=100\GeV$. The dijet mass $\Mjj$ is calculated from the
corrected four-momentum vectors of the two jets with the highest \pt
(leading jets). The relative dijet-mass resolution, estimated from the
simulation, ranges from 7\% at $\Mjj=0.2\TeV$ to $3\%$ at
$\Mjj=3\TeV$.
\section{Data samples and event selection}

The data samples used for this measurement were collected with
single-jet high-level triggers (HLT)~\cite{HLT} that require at least
one jet in the event to have $\pt > 60$, 110, 190, 240, or
370\GeV, respectively, in corrected jet transverse momentum. The
online jet reconstruction uses only calorimetric information and the
resulting HLT jets typically have worse energy resolution than the
offline particle-flow jets. The lower-\pt triggers were prescaled and
the corresponding integrated luminosity of each trigger sample,
$\mathcal{L}_{\text{eff}}$, is listed in Table~\ref{tab:lumi}. In the
offline analysis, events are required to have at least one well
reconstructed proton-proton interaction vertex~\cite{TRK-10-005}. In
order to suppress nonphysical jets, \ie, jets resulting from noise in
the electromagnetic and/or hadronic calorimeters, the jets are
required to satisfy the following identification criteria. Each jet
should contain at least two particles, one of which is a charged
hadron, and the jet energy fraction carried by neutral hadrons and
photons should be less than 90\%. These criteria have an efficiency
of greater than 99\% for physical jets, while the probability for a 
nonphysical jet to pass the criteria is less than $10^{-6}$.

\begin{table}[htbH]
  \begin{center}
    \topcaption{The integrated luminosity for each of the data samples.
      \label{tab:lumi}}
    \begin{scotch}{cccccc}
      Min. jet trigger \pt& 60 & 110 & 190 & 240 & 370 \\
       (\GeVns{}) &&&&&\\
      Jet trigger name & Jet60 & Jet110 & Jet190 & Jet240 & Jet370 \\
      $\mathcal{L}_{\text{eff}}$ $(\!\pbinv)$ & 0.41 & 7.3 & 152 & 512 & 4980 \\
    \end{scotch}
  \end{center}
\end{table}

\begin{figure}[ht]
  \centering
  \includegraphics[width=0.45\textwidth]{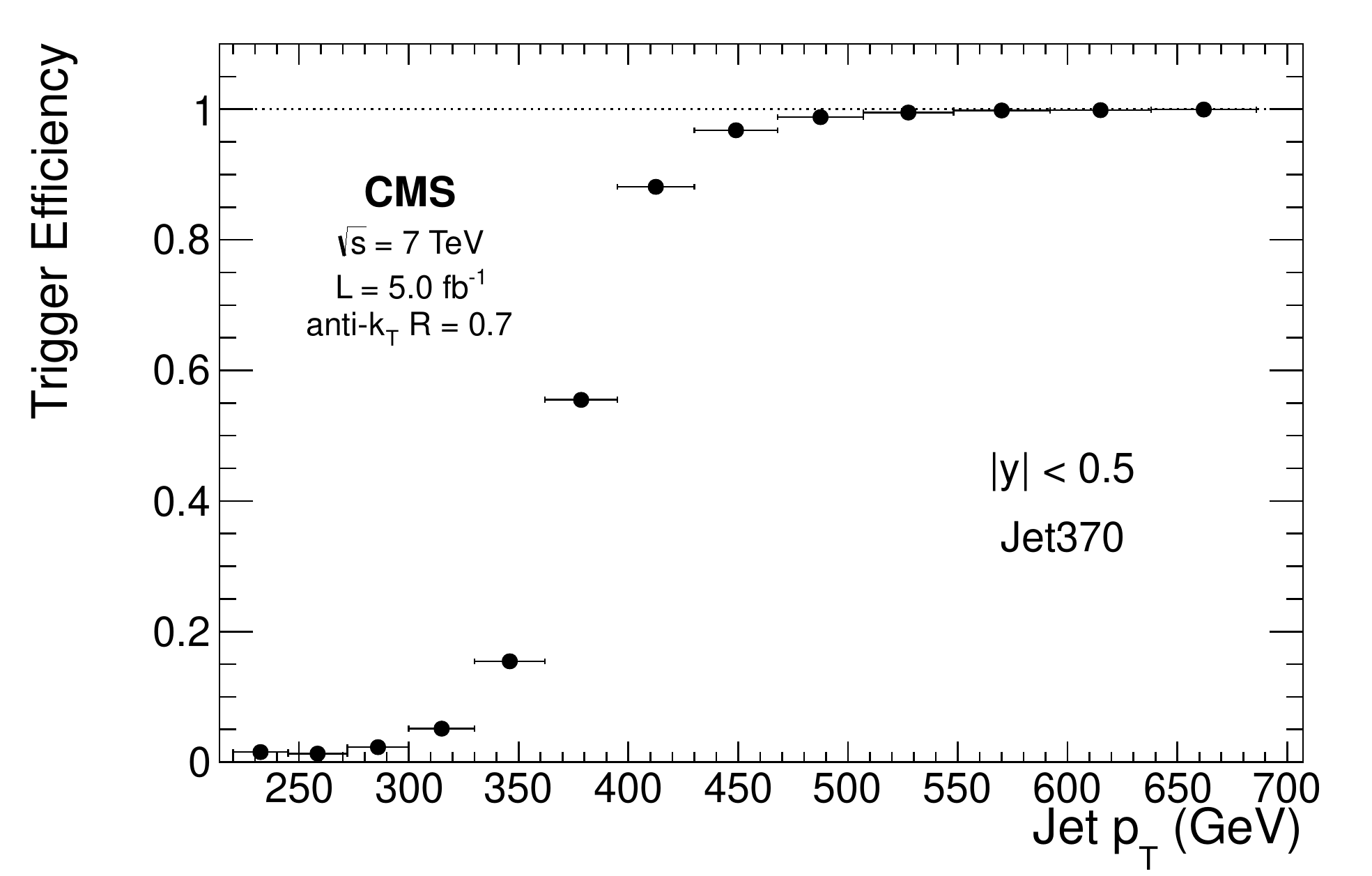}
  \includegraphics[width=0.45\textwidth]{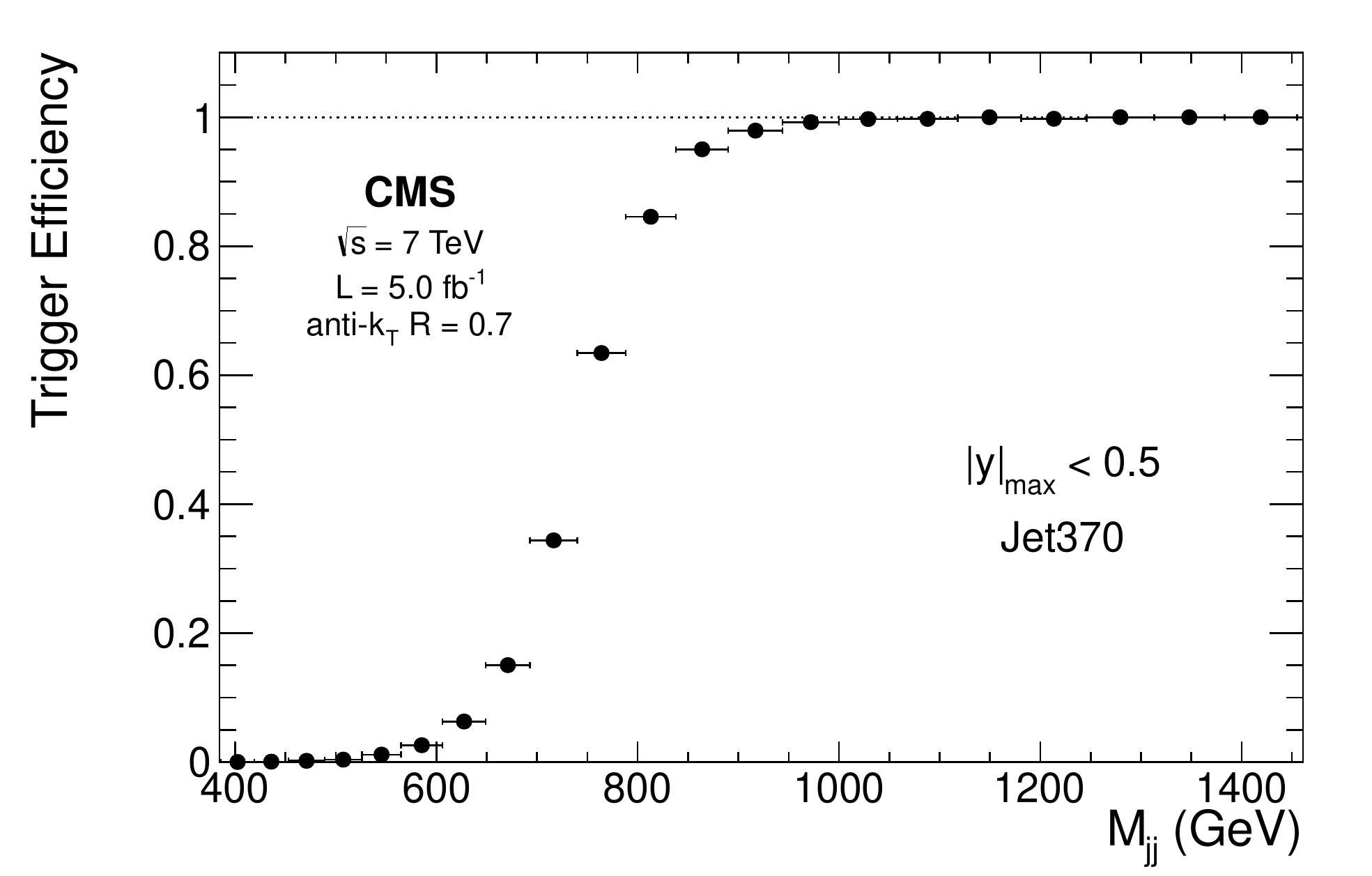}
  \caption{Trigger efficiency as a function of the jet \pt (\cmsLeft) and
    dijet mass $\Mjj$ (\cmsRight) for the 370\GeV single-jet
    trigger and for the central rapidity bins.}
  \label{fig_trig}
\end{figure}

The inclusive single-jet cross section measurements are made in five
rapidity regions of size $\Delta|y|=0.5$ over the range 0.0--2.5. Jets
that do not satisfy the jet identification criteria are discarded and
events are required to contain at least one jet that satisfies these
criteria.  In order to avoid any trigger bias, the jets are
additionally required to have $\pt>110$, 200, 300, 360, and
510\GeV for the five single-jet HLT triggers used, respectively.
Figure~\ref{fig_trig} (\cmsLeft) shows the trigger efficiency as a
function of the jet \pt, for the central rapidity bin $|y|<0.5$ and
for the highest trigger threshold. The efficiency of each trigger path
has been measured using events collected with a lower threshold
single-jet trigger and confirmed with events collected with
single-muon triggers.

\begin{figure}[ht]
  \centering
  \includegraphics[width=0.45\textwidth]{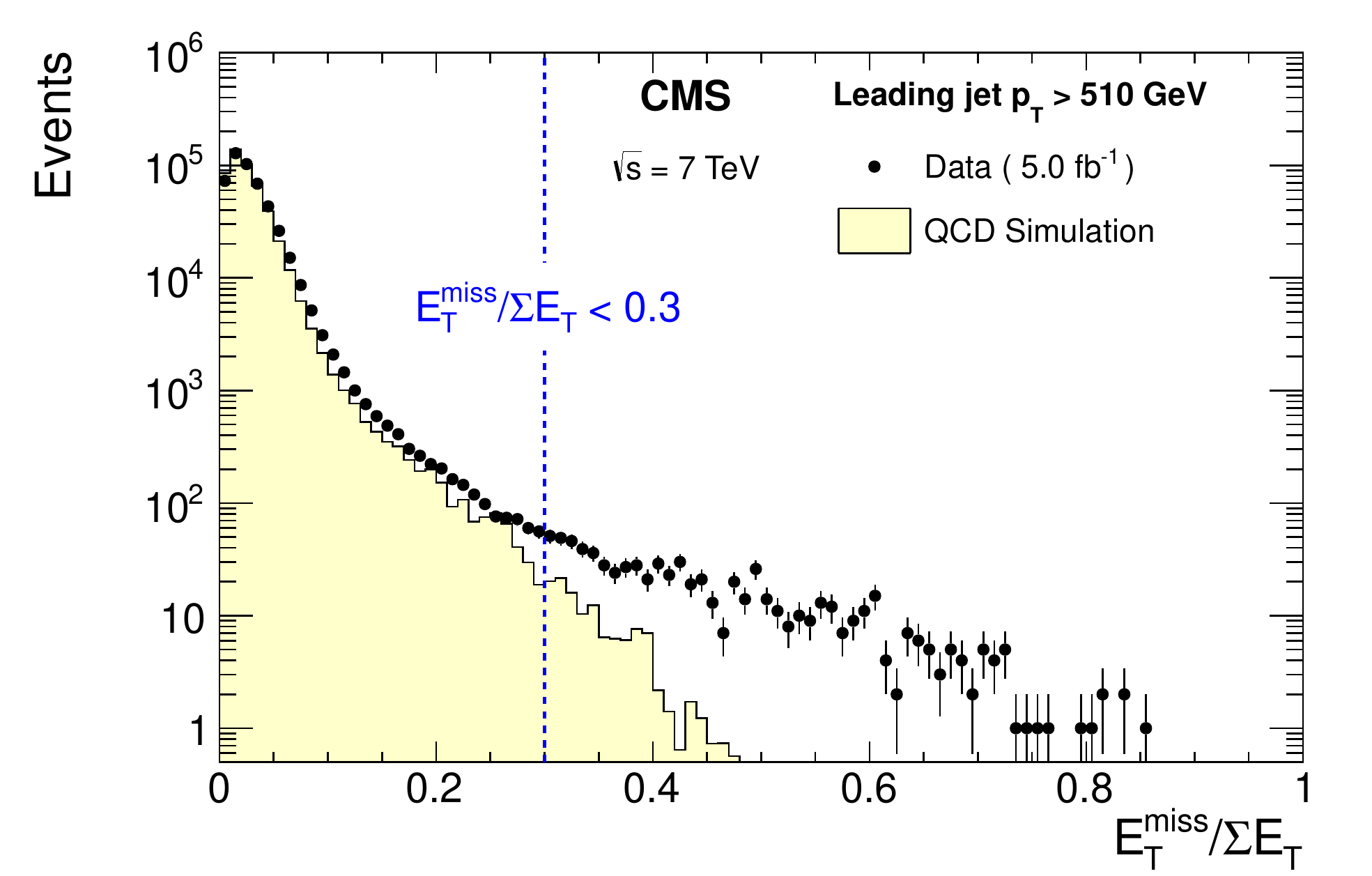}
  \includegraphics[width=0.45\textwidth]{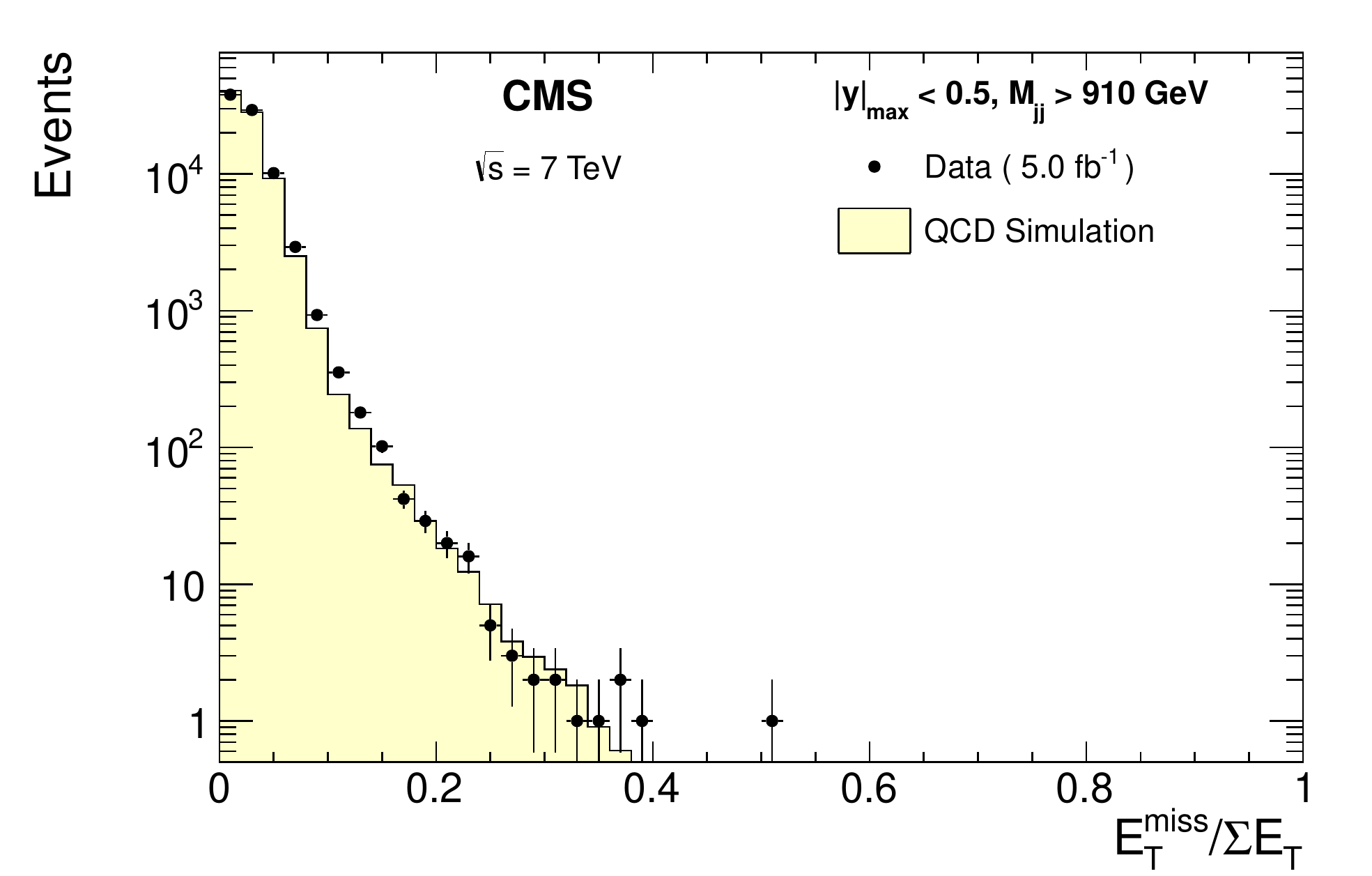}
  \caption{Distribution of $\ETmiss/\Sigma\,\ET$ for data
    events (black points) and simulated QCD events (continuous line)
    with at least one jet with $\pt>510\GeV$ (\cmsLeft) and for dijet
    events with $|y|_\text{max}<2.5$ and $\Mjj>910\GeV$
    (\cmsRight), collected with the 370\GeV single-jet trigger. The
    distribution from the inclusive-jet selection is shown before the
    offline selection $\ETmiss/\Sigma\,\ET<0.3$.  The larger
    tail in the data is caused by other processes with true
    \ETmiss (such as \cPZ+jet(s), where the \cPZ\ boson decays to neutrinos,
    and \PW+jet(s), where the \PW\ boson decays to leptons).}
  \label{fig_met}
\end{figure}

Background events due to instrumental noise, beam halo effects, or
proton-proton collisions with leptons in the final state that might survive the jet
identification criteria are further suppressed by requiring
$\ETmiss/\Sigma\,\ET<0.3$. Hard QCD processes do not generate
true $\ETmiss$ and because of the good energy resolution
the measured values of \ETmiss in such events are small compared to the total transverse energy.
Hence, the distribution of the variable $\ETmiss/\Sigma\,\ET$ peaks close
to zero for QCD events, while some background events give
larger values. Figure~\ref{fig_met} (left) shows a typical distribution of the variable $\ETmiss/\Sigma\,\ET$ for events
with at least one jet with $\pt>510\GeV$, collected with the 370\GeV single-jet trigger.
The data points exceed the QCD predictions at large values of \ETmiss. This tail is dominated by processes such as \cPZ+jet(s),
where the \cPZ\ boson decays to neutrinos, and \PW+jet(s), where the \PW\ boson
decays to leptons. No events from instrumental and beam halo effects were found to survive the
jet identification criteria, and therefore the additional
$\ETmiss/\Sigma\,\ET$ selection only removes events from processes that produce true $\ETmiss$.

For the dijet measurement, at least two jets with ${\pt}_1>60\GeV$ and
${\pt}_2>30\GeV$ and satisfying the tight identification criteria are
required. If either of the two leading jets fails the identification
criteria, the event is discarded. The dijet measurement is performed
in five rapidity regions of size $\Delta|y_\text{max}|=0.5$, defined by the maximum absolute rapidity
$y_\text{max}=\sign\left(|\max\left(y_1,y_2\right)|-|\min\left(y_1,y_2\right)|\right)\cdot\max\left(|y_1|,|y_2|\right)$ of the two leading
jets in the event. The use of the variable $|y|_\text{max}\equiv|y_\text{max}|$ divides
the phase space of the dijet system into exclusive rapidity bins,
which correspond to different scattering angles in the center-of-mass
frame.

Low values of $|y|_\text{max}$ probe $s$-channel scattering at large angles,
while large values of $|y|_\text{max}$ probe $t$-channel scattering at small angles.
For each rapidity bin the trigger
efficiency is expressed as a function of $\Mjj$, and the events
are required to satisfy a minimum mass threshold, which increases with
$|y|_\text{max}$.  Figure~\ref{fig_trig} (\cmsRight) shows the trigger
efficiency for the central rapidity bin and for the
highest trigger threshold. Because of the dijet topology, no further cut
is needed on the $\ETmiss/\Sigma\,\ET$ variable, as can be seen
in Fig.~\ref{fig_met} (\cmsRight).

\section{Measurement of the differential jet and dijet cross
  sections}\label{sec:Exp}

In this section the reconstruction of the jet transverse momentum and
dijet mass spectra from the different samples is presented. Then the
unfolding procedure, which translates the reconstructed spectra into
true spectra, is described. Finally, the experimental uncertainties
related to the measurements are described and discussed.

\subsection{Determination of transverse momentum and dijet mass
  spectra}

\begin{figure}[ht]
  \centering
  \includegraphics[width=0.45\textwidth]{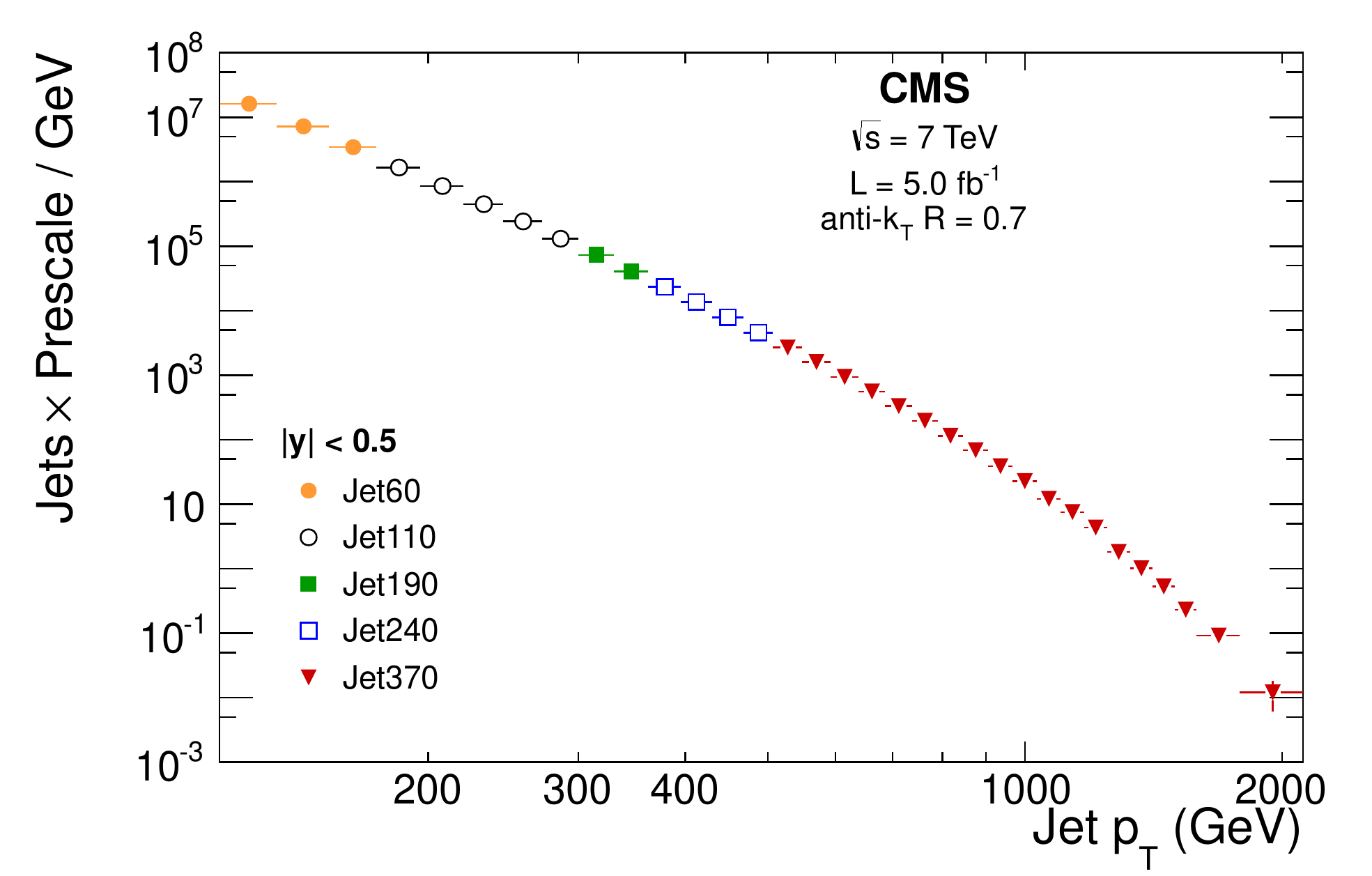}
  \includegraphics[width=0.45\textwidth]{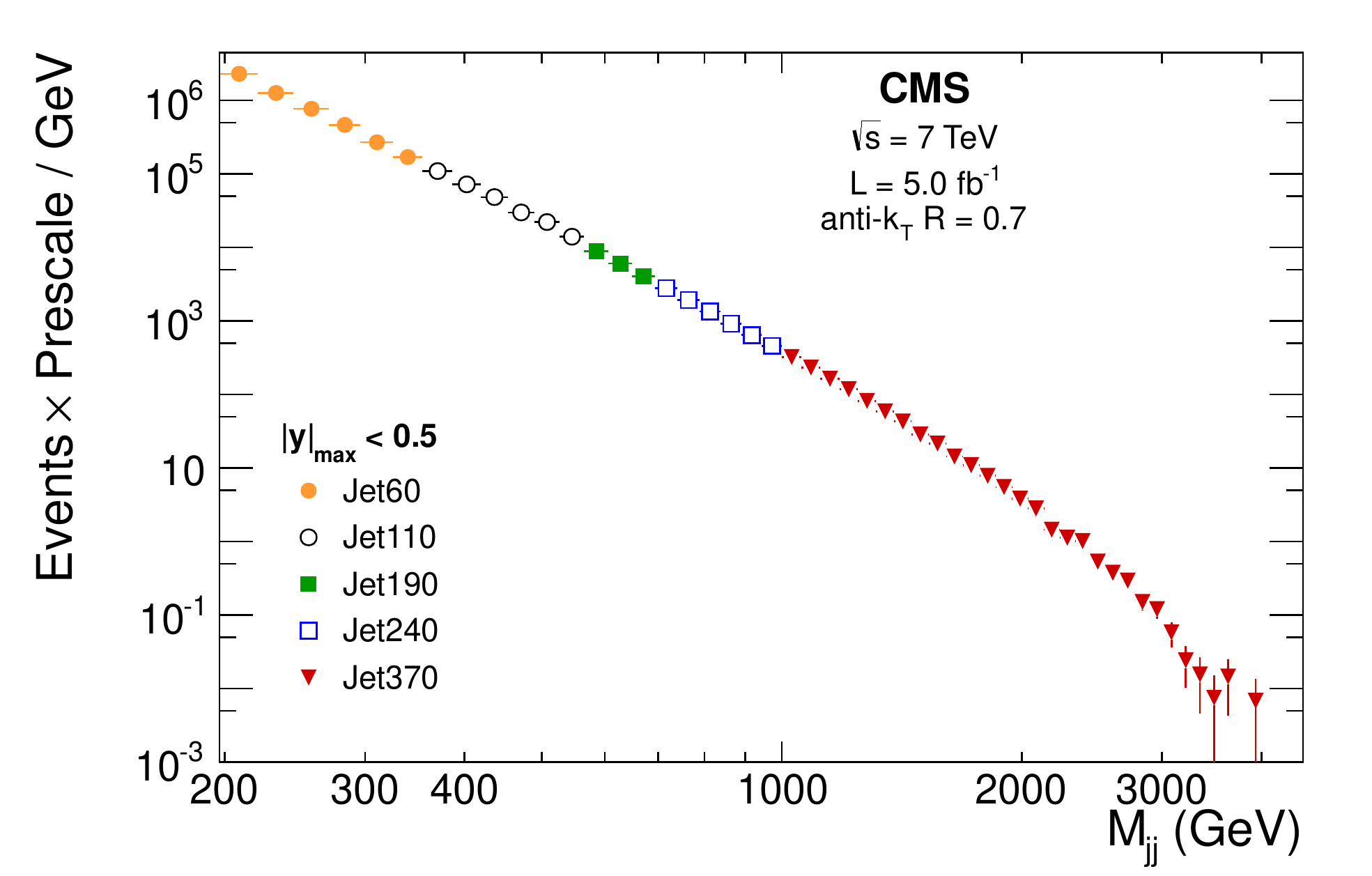}
  \caption{Spectrum construction from individual trigger paths. \cmsLLeft:
    inclusive jet \pt spectrum for $|y|<0.5$. \cmsRRight: dijet mass
    spectrum for $|y|_\text{max}<0.5$. The different markers indicate
    different trigger paths.}
  \label{fig:spectra_construction}
\end{figure}

The jet \pt (dijet invariant mass) spectrum is obtained by populating
each bin with the number of jets (events) collected using the highest
threshold trigger which gives more than 99\% trigger efficiency.
Then, the yields from each trigger path are scaled according to the
corresponding prescale value for this path (effective luminosity), as
shown in Table~\ref{tab:lumi}. Figure~\ref{fig:spectra_construction}
shows the reconstructed spectra for inclusive jets (\cmsLeft) and dijets
(\cmsRight), in the central rapidity bin, decomposed into the five
contributing trigger paths.

The observed inclusive jet yields are transformed into double-differential cross sections as
follows:
\begin{equation}\label{master_inclusive}
  \frac{\rd^2\sigma}{\rd\pt\,\rd y} =
  \frac{1}{\epsilon\cdot\mathcal{L}_{\text{eff}}}
  \frac{N_\text{jets}}{\Delta\pt\,\left(2\cdot\Delta |y|\right)},
\end{equation}
where $N_\text{jets}$ is the number of jets in the bin,
$\mathcal{L}_{\text{eff}}$ is the integrated luminosity of the data
sample from which the events are taken, $\epsilon$ is the product of
the trigger and event selection efficiencies, both of which are
greater than 99\%, and $\Delta \pt$ and $\Delta|y|$ are the transverse momentum
and rapidity bin widths, respectively. The width of the \pt bins is
proportional to the \pt resolution and so increases with \pt. The
statistical uncertainty assigned to each \pt bin takes into account
the number of independent events that contribute at least one
jet in the bin. The largest fraction (more than 90\%) of the observed
jets in each \pt bin originate from different events, however a small fraction of
events contribute more than one jet. Such
events are typically back-to-back dijet events, so closely balanced in
\pt that both jets end up in the same $|y|$ and \pt bin.

The statistical uncertainty in the number of jets in a bin is
$e_\text{stat}=\sqrt{(4-3f)/(2-f)}\cdot\sqrt{N_\text{jets}}$, where
$f=N_1/N_\text{ev}$ is the fraction of events that
contribute one jet in the given bin.
The formula is valid under the assumption that the number of events that
contribute more than two jets in each bin is negligible, which has
been verified for the current measurement.

The observed dijet yields are transformed into double-differential cross sections as
follows:
\begin{equation}\label{master_formula}
  \frac{\rd^2\sigma}{\text{d}\Mjj\,\rd y_\text{max}} =
  \frac{1}{\epsilon\cdot\mathcal{L}_{\text{eff}}}\cdot
  \frac{N}{\Delta \Mjj\,\left(2\cdot\Delta|y|_\text{max}\right)},
\end{equation}
where $\Delta \Mjj$ and $\Delta|y|_\text{max}$ are the mass and
rapidity bin widths, respectively.  The size of the dijet mass bins is
approximately equal to or larger than the mass resolution at the bin
center, while the bins at the edge of the spectrum have been merged to
assure a minimal number of events in each bin.

\subsection{Unfolding}

\begin{figure}[ht]
  \centering
  \includegraphics[width=0.45\textwidth]{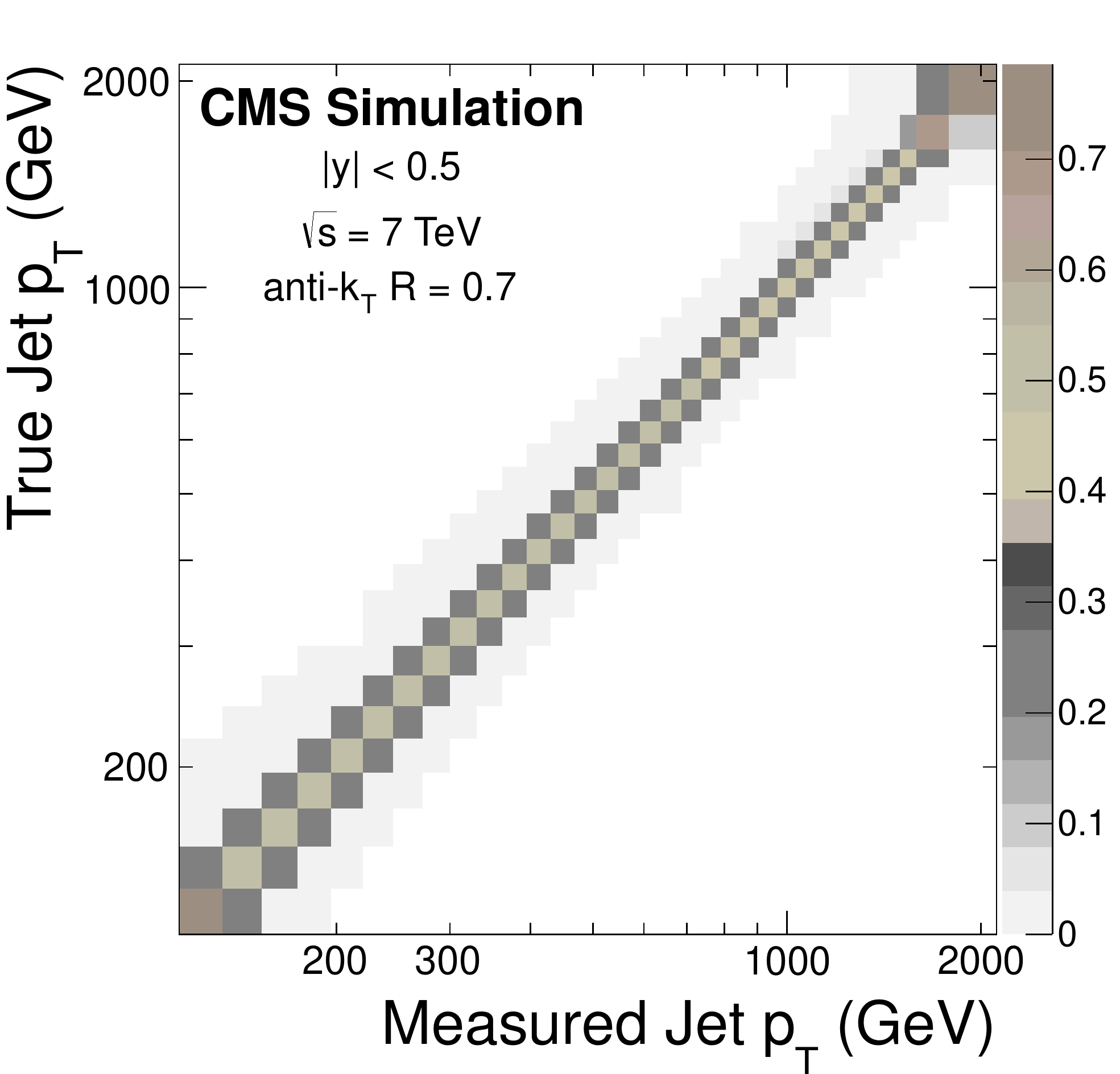}
  \includegraphics[width=0.45\textwidth]{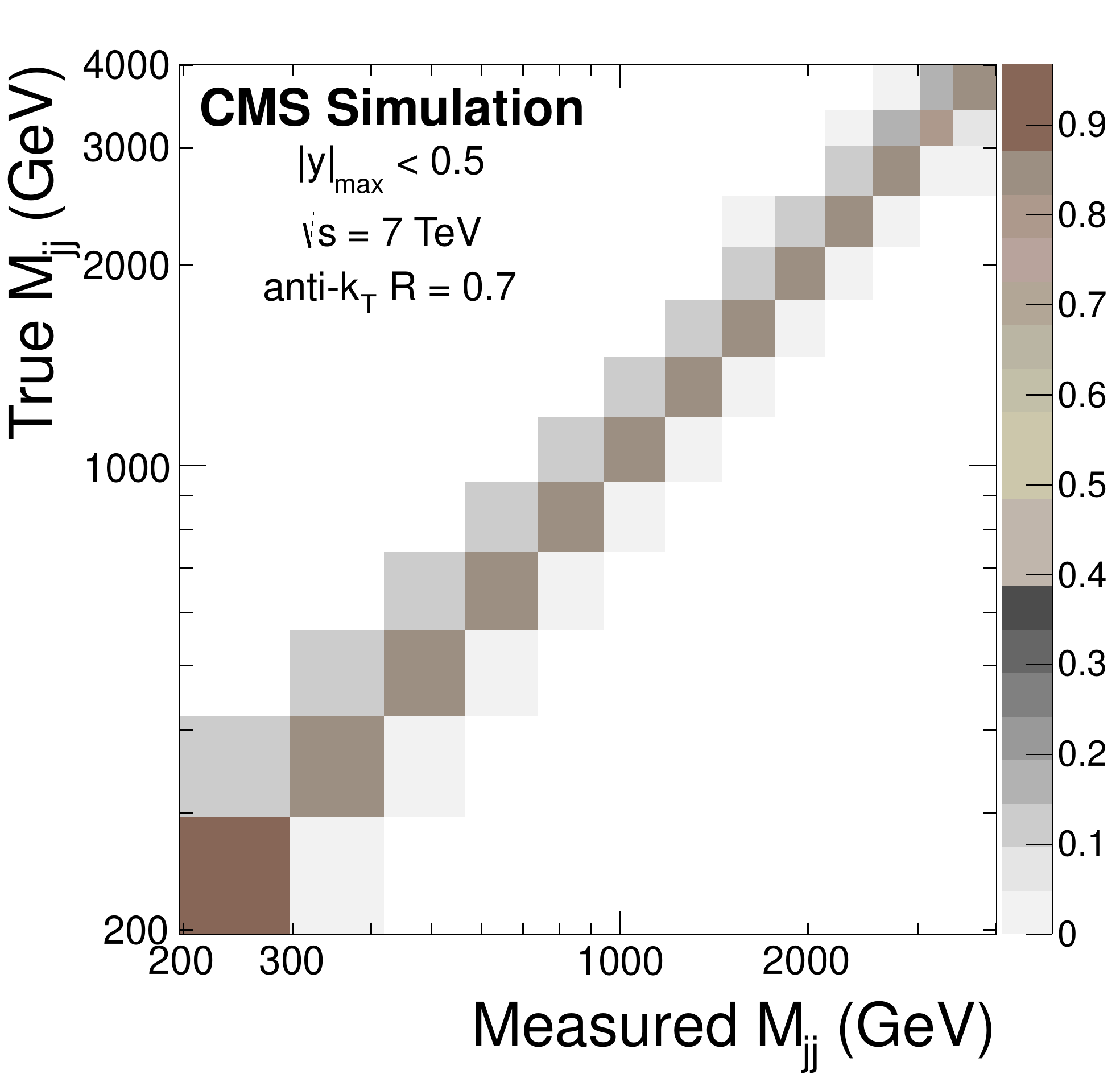}
  \caption{Response matrices for the inclusive jet \pt spectrum (\cmsLeft)
    and the dijet mass spectrum (\cmsRight) in the central rapidity bins.}
  \label{fig_responsemat}
\end{figure}

Because of the detector resolution and the steeply falling
spectra, the measured differential cross sections are smeared with
respect to the particle-level cross sections. Each \pt and mass bin
contains events that have migrated in from neighboring bins and is
missing events that have migrated out. For a steeply falling
spectrum more events migrate into a bin than out. In order to allow for a direct comparison of
experimental measurements with corresponding results from other
experiments and with QCD predictions, the spectra are unfolded in order to correct for detector effects.
The response matrix is obtained from
the detector simulation and corrected for the measured differences in the
resolution between data and simulation~\cite{JES}.
Figure~\ref{fig_responsemat} shows the response matrices for the jet
\pt (\cmsLeft) and the dijet mass (\cmsRight) in the central rapidity bins.
The unfolding is done with the RooUnfold package~\cite{RooUnfold}
using the D'Agostini method~\cite{Bayes}.

\subsection{Experimental uncertainties}

\begin{figure*}[htbp]
  \centering
  \includegraphics[width=0.40\textwidth]{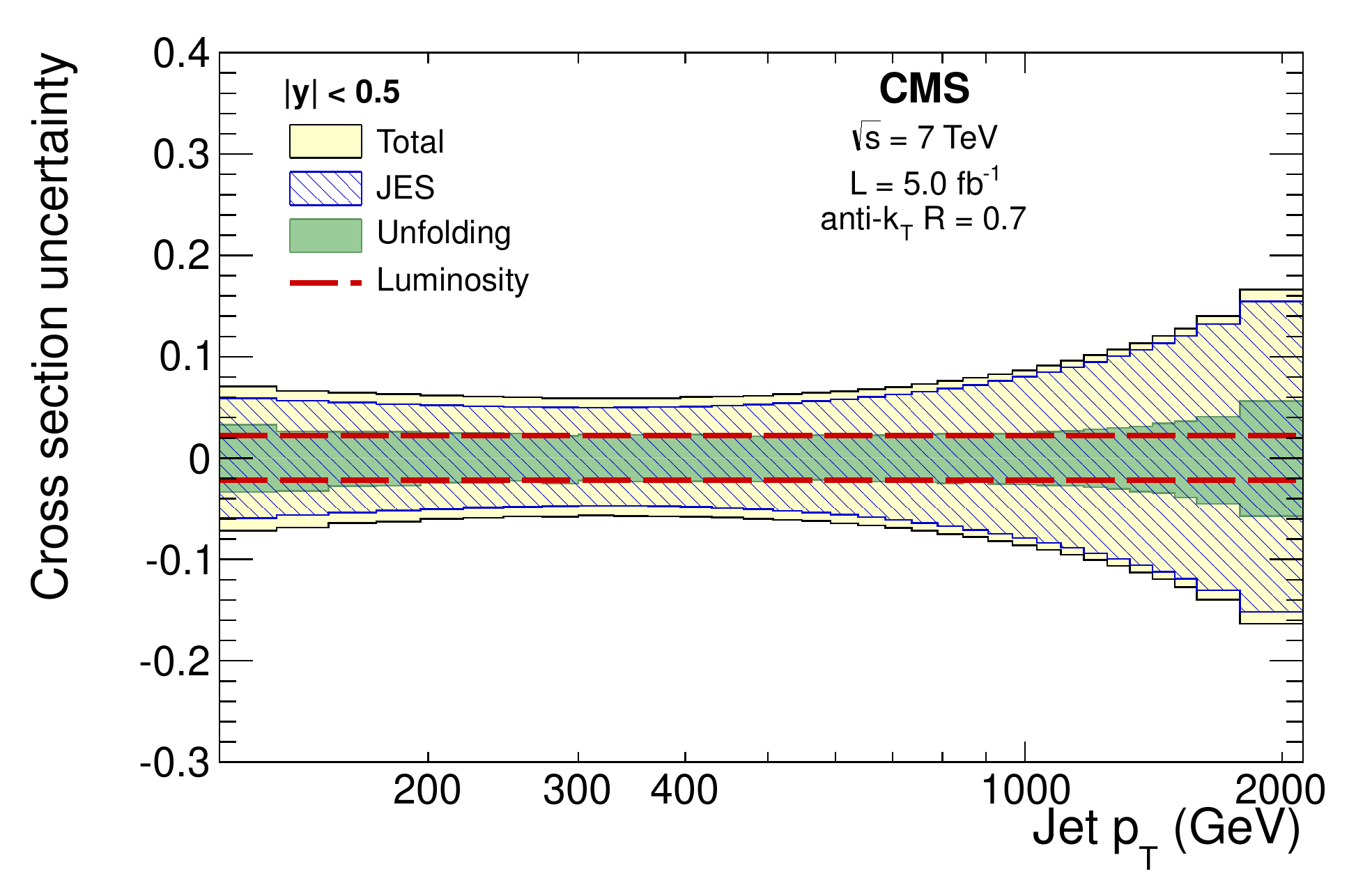}
  \includegraphics[width=0.40\textwidth]{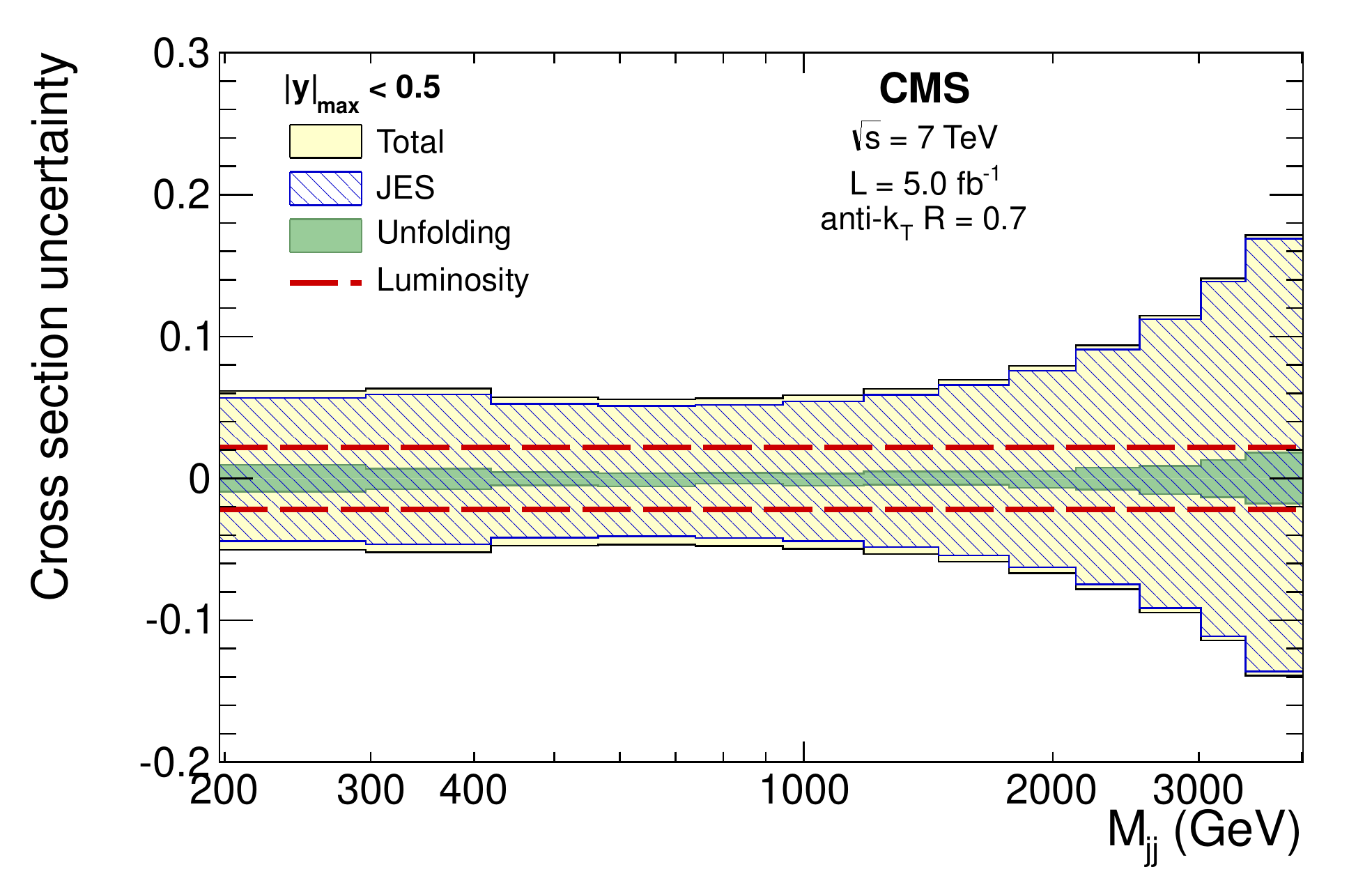}
  \includegraphics[width=0.40\textwidth]{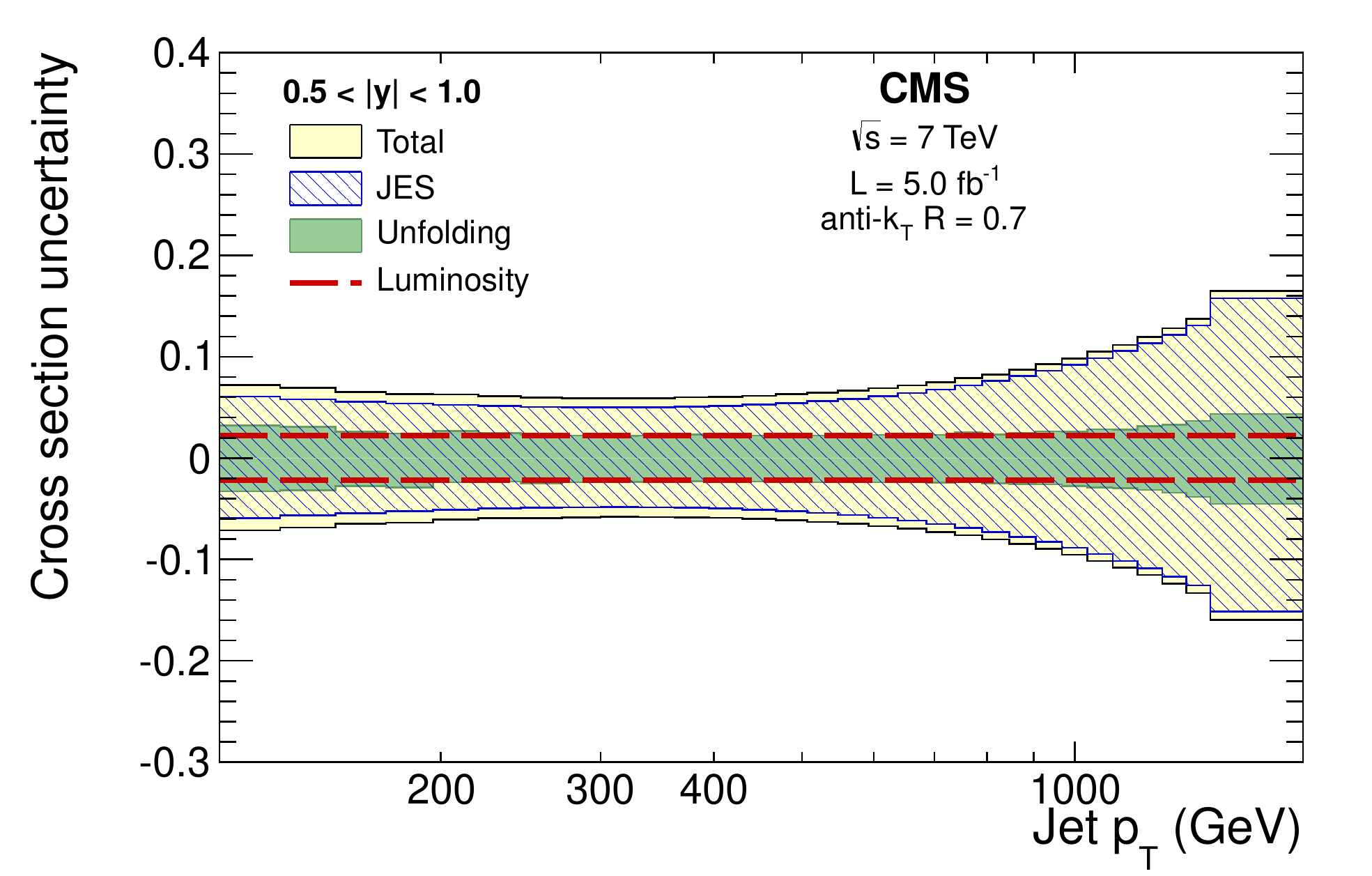}
  \includegraphics[width=0.40\textwidth]{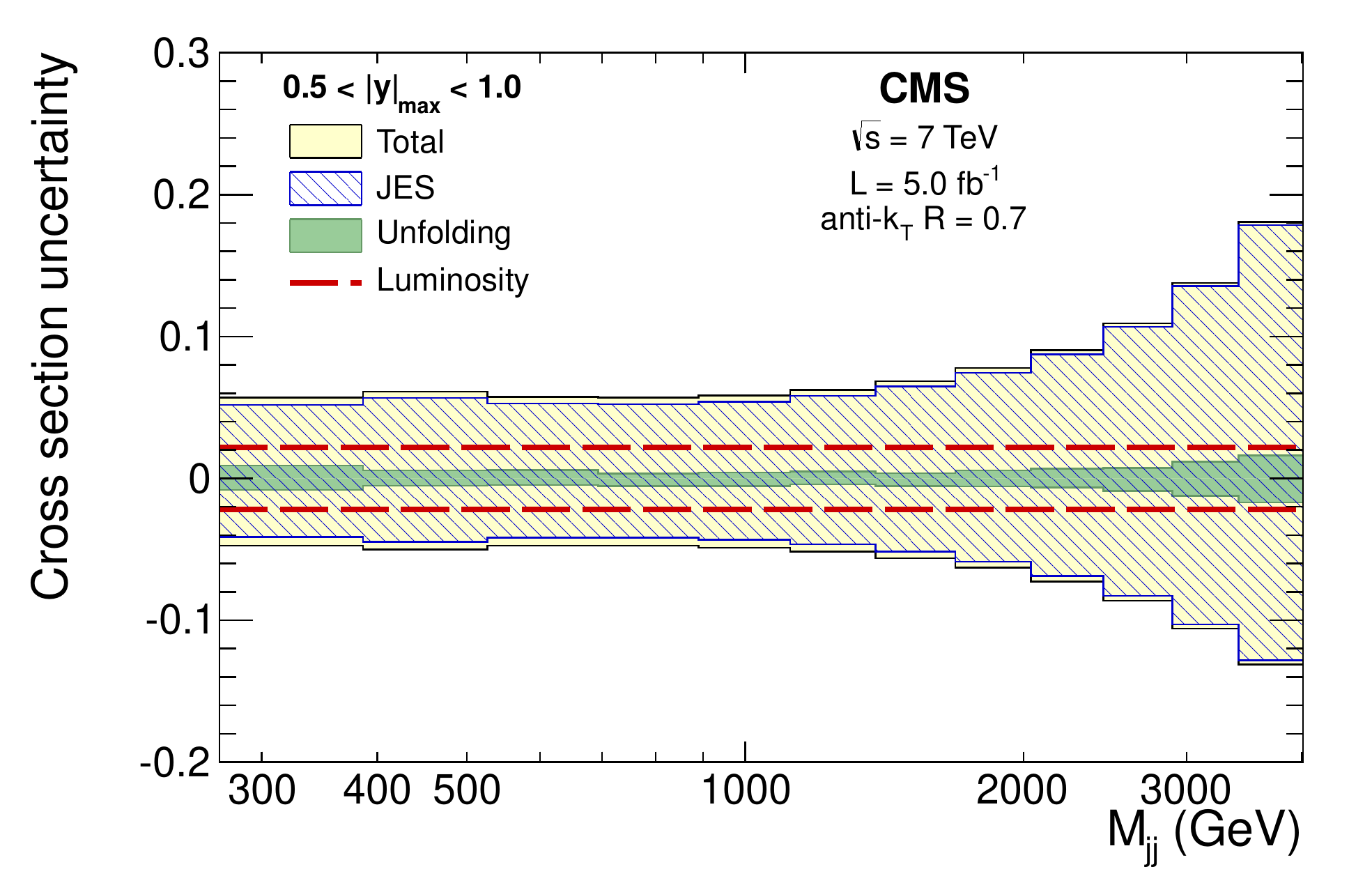}
  \includegraphics[width=0.40\textwidth]{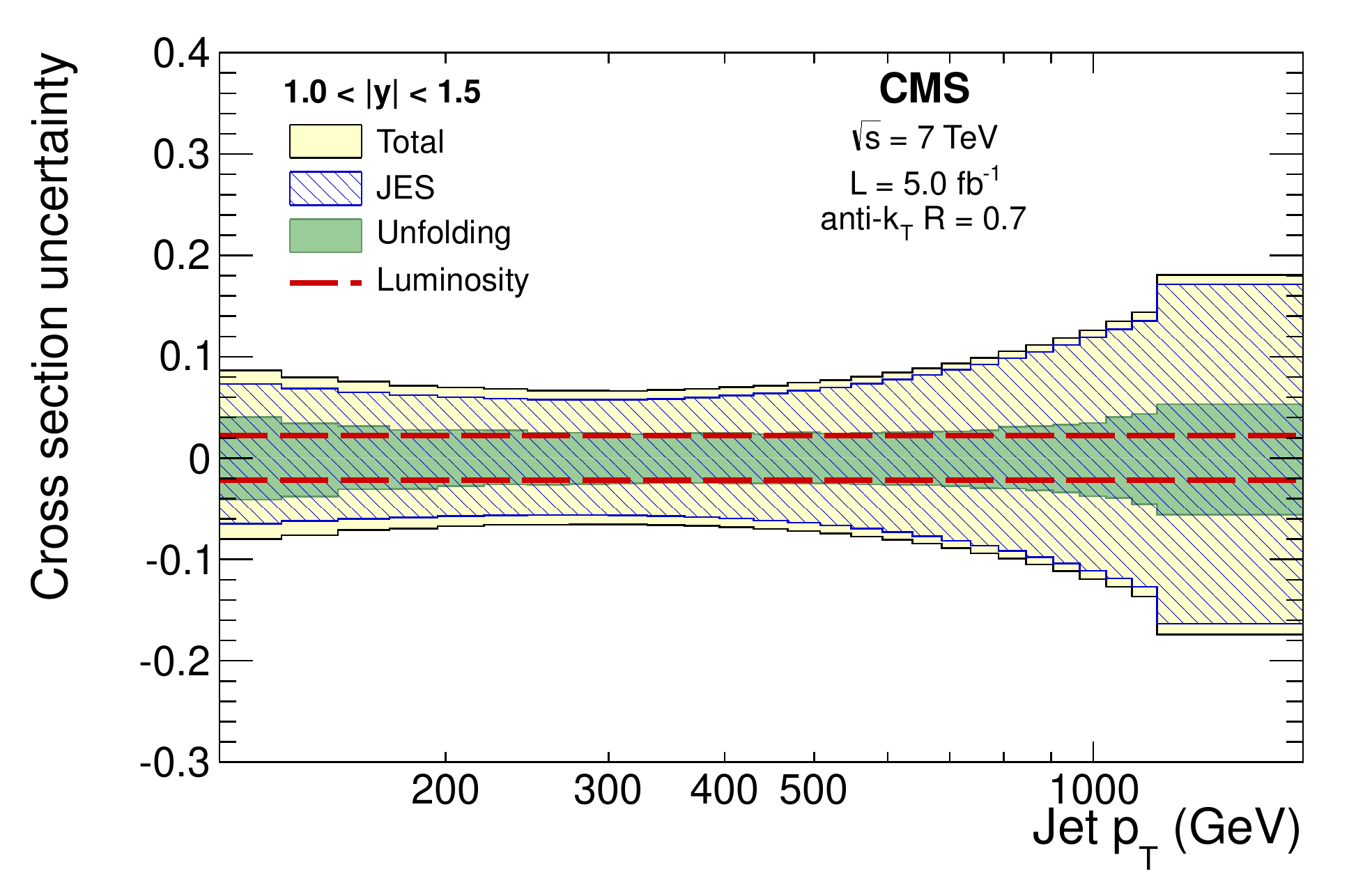}
  \includegraphics[width=0.40\textwidth]{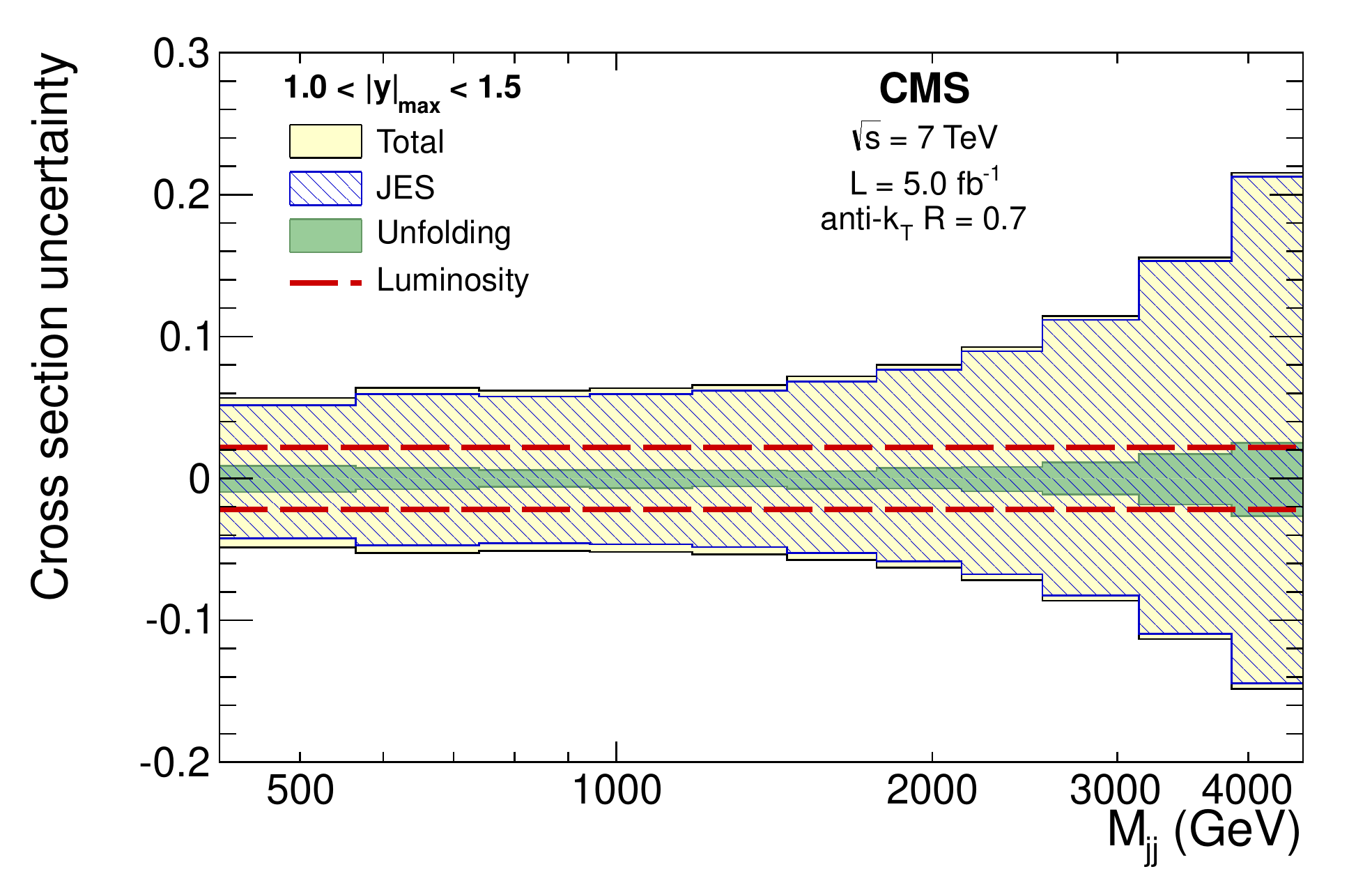}
  \includegraphics[width=0.40\textwidth]{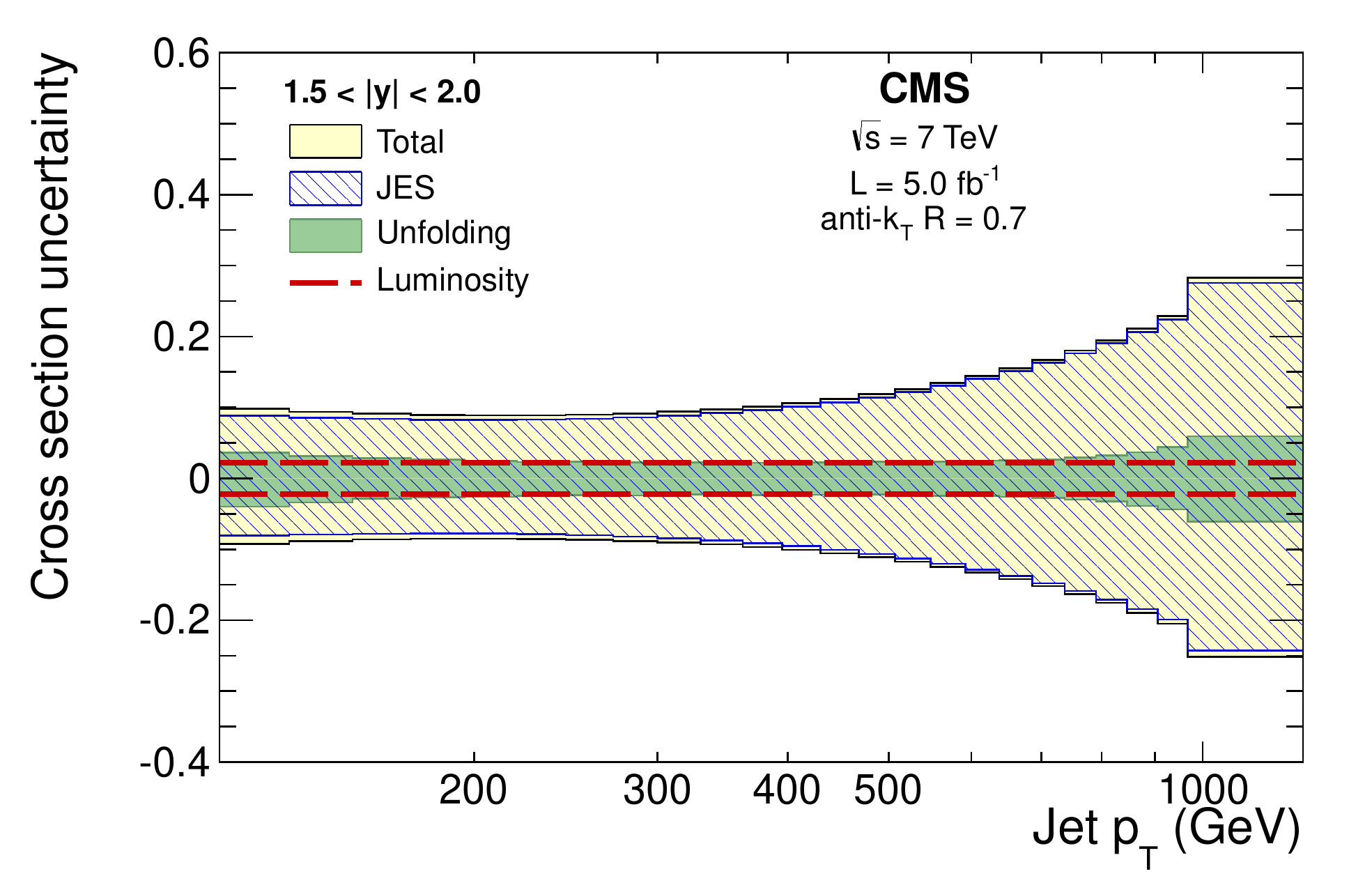}
  \includegraphics[width=0.40\textwidth]{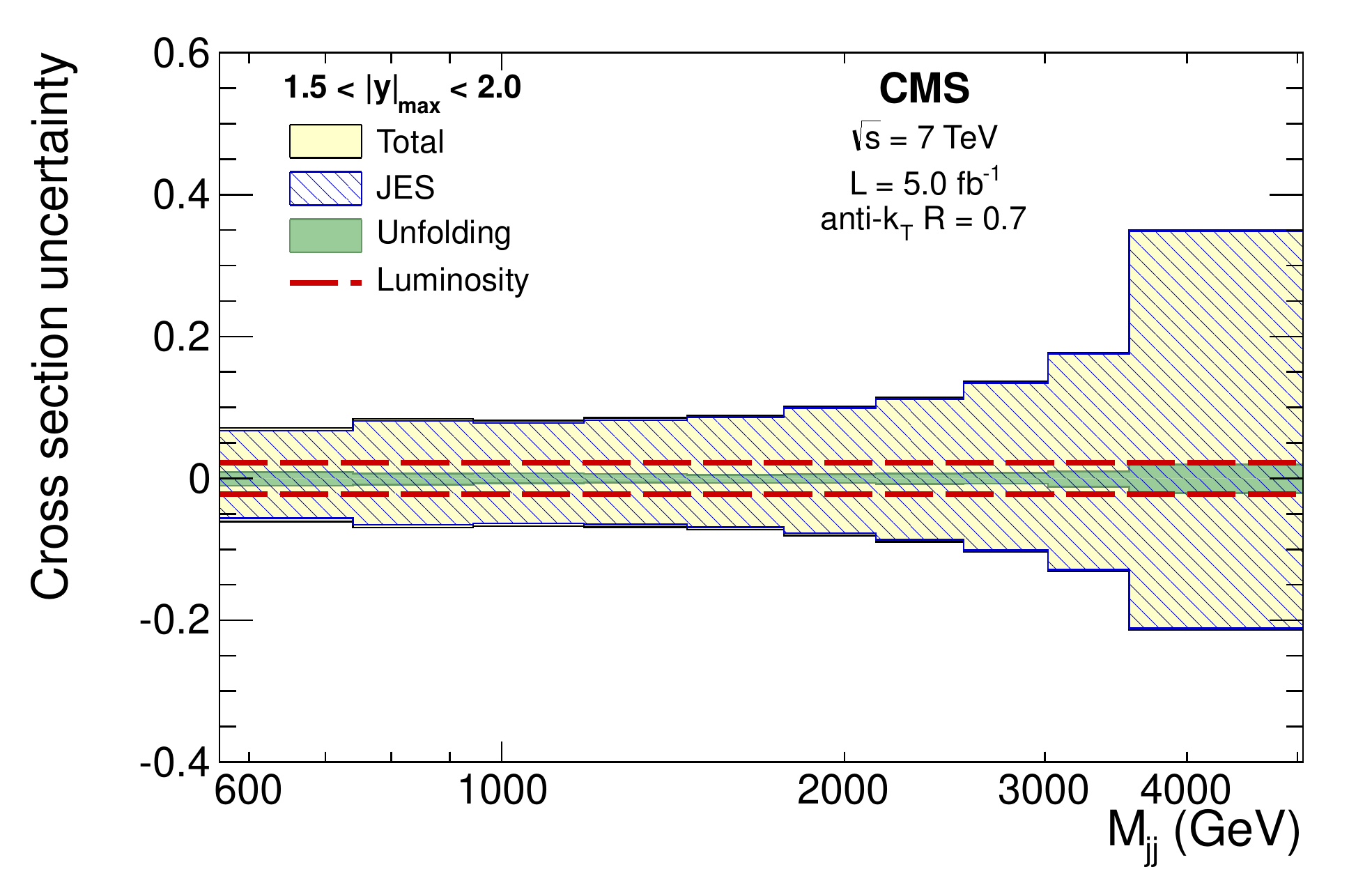}
  \includegraphics[width=0.40\textwidth]{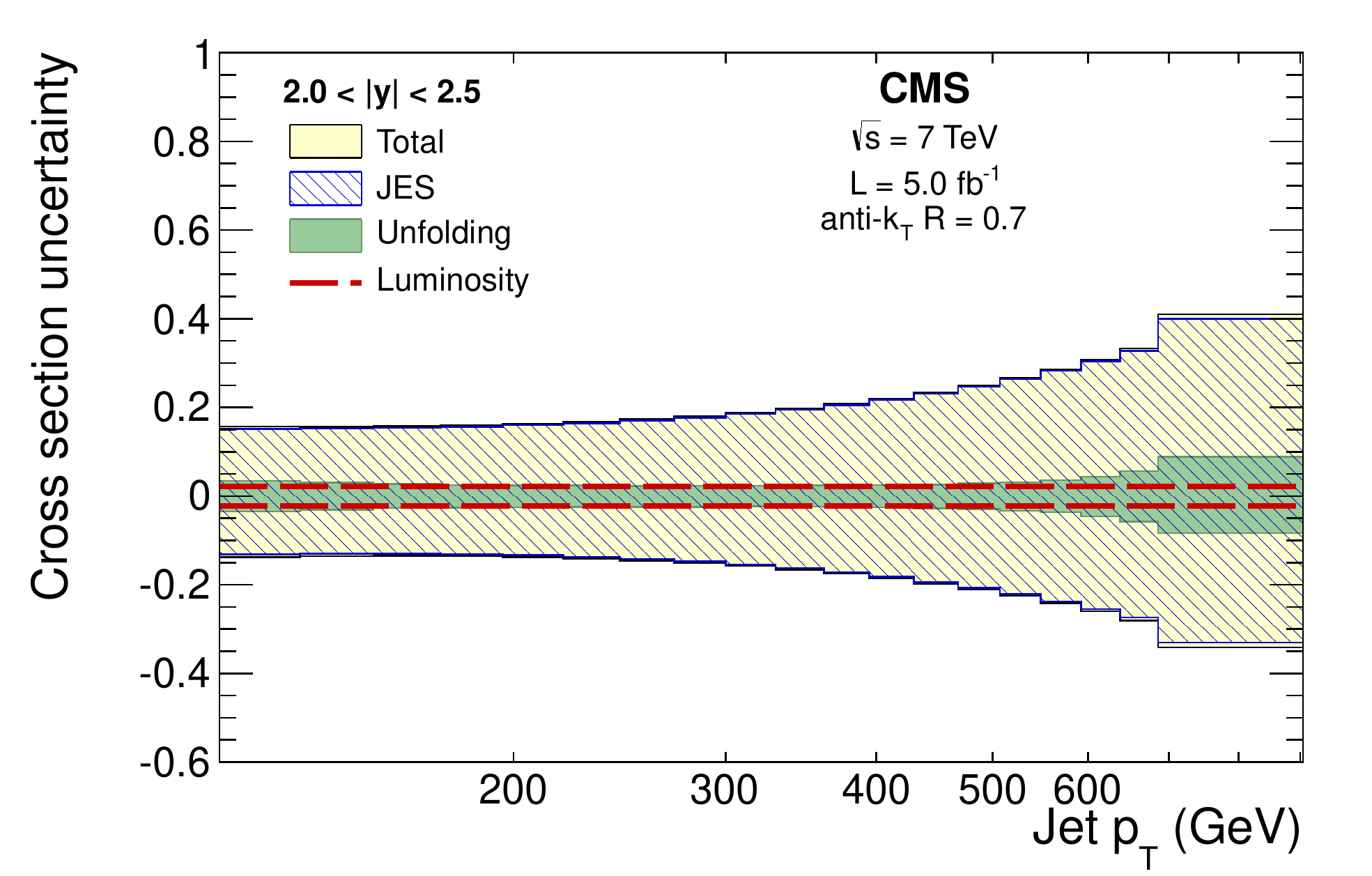}
  \includegraphics[width=0.40\textwidth]{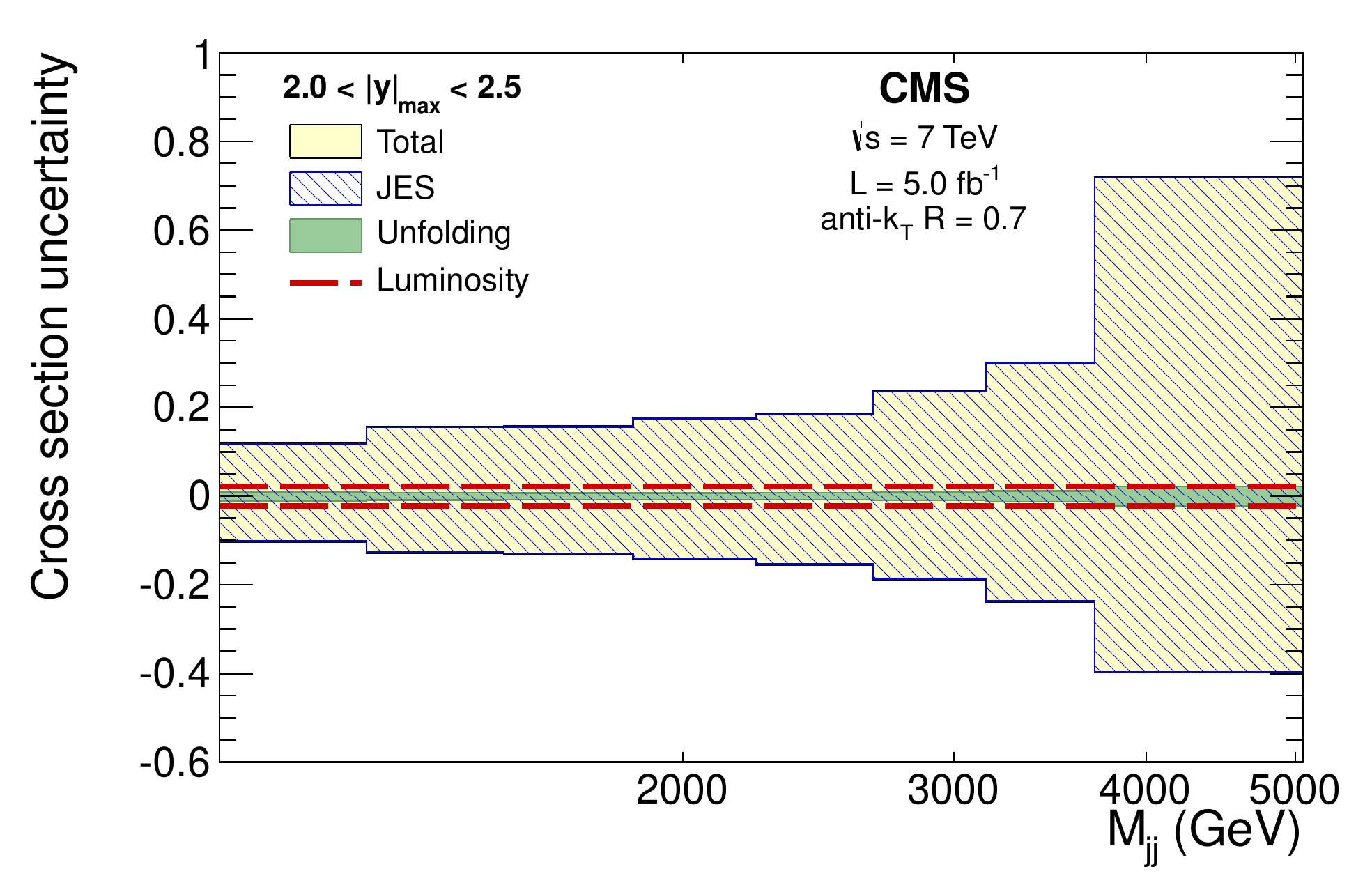}
  \caption{Effect of the relative experimental uncertainties for the
    inclusive jet (left column) and dijet (right column) cross section
    measurements, and for all five $|y|$ and
    $|y|_\text{max}$ bins, respectively. The upward and downward uncertainties are estimated
    separately.}
  \label{fig_EXPunc}
\end{figure*}

The dominant experimental uncertainties are related to the jet energy
scale (JES), the luminosity, and the jet \pt resolution. Other sources
of systematic uncertainty, such as the jet angular resolution, are
negligible. The agreement of the results for positive
and negative rapidities has also been confirmed.
Figure~\ref{fig_EXPunc} shows the effects of the experimental
uncertainties in all rapidity bins for the cross section measurements.
For rapidities up to $|y| = 1.5$ the total uncertainty of both cross
sections ranges from 5\% at low \pt or $\Mjj$ to $20\%$ at
high \pt or $\Mjj$, respectively.  For higher rapidities the
total uncertainty increases to 10--30\% in both cases, with the
exception of the highest dijet mass bin in the outer rapidity region
of $2.0 < |y|_\text{max} < 2.5$, where the uncertainty is
substantially larger.  A discussion of the individual contributions to
the uncertainty follows.

\subsubsection{Jet energy scale (JES) uncertainty}

The jet energy scale is the dominant source of systematic uncertainty.
Because of the steep slope of the \pt spectrum, a small uncertainty in the
\pt scale translates into a large uncertainty in the cross section for
a given value of \pt. The jet energy scale uncertainty is dependent on
\pt and $\eta$ and has been estimated to be 2.0--2.5\%~\cite{JES2011}.
The individual, uncorrelated contributions to the JES uncertainty have
been estimated and are discussed below.

The JES uncertainty sources account for the \pt and $\eta$ dependence
of the JES within the total uncertainty.  For the phase space of jets
considered here, 16 mutually uncorrelated sources contribute to the
total uncertainty, where each such source represents a signed
$1\sigma$ variation from a given systematic effect for each point in
$(\pt,\eta)$. Summing up separately the positive or negative
variations of the sources in quadrature will reproduce the total
upward and downward JES uncertainties at each point. 
The uncertainties from all 16 independent sources are included in \suppMaterial\
submitted with this paper and in the HEPDATA record 
for this  paper; the cross section measurements and other details are also 
tabulated therein.

The uncertainty sources are divided into four broad categories: pileup
effects, relative calibration of jet energy scale versus $\eta$,
absolute energy scale including \pt dependence, and differences in
quark- and gluon-initiated jets. The first category, containing pileup
effects, has relatively little impact on the analyses presented in this
paper.

The second category, containing $\eta$-dependent effects,
parameterizes the possible relative variations in JES, which for the
dijet and inclusive jet analyses lead to correlations between
rapidity bins. In principle these effects could also have a \pt
dependence, but systematic studies on data and Monte Carlo (MC) events indicate that the
\pt and $\eta$ dependence of the uncertainties factorize to a good
approximation.

The third category deals with the uncertainty in the absolute energy scale
and its \pt dependence and is the most relevant one for these
analyses.  The photon+jet and \cPZ+jet events only constrain the JES
directly in a limited jet \pt range of about 30--600\GeV, and the
response at higher (and lower) \pt is estimated by MC simulation. The
\pt-dependent uncertainty arising from modeling of the underlying event
and jet fragmentation is obtained by comparing predictions from \PYTHIA{6} and \HERWIG{++}. Most studies show that both generators
agree with the data with differences comparable to those seen between
data and MC\@. The uncertainty arising from the calorimeter response
to single hadrons is estimated by varying the response
parameterization by ${\pm}3$\%  around the central value.
The final uncertainty arises from differences in the JES for quark- and
gluon-initiated jets and is determined from MC studies.

\subsubsection{Luminosity uncertainty}

The luminosity uncertainty is estimated to be
2.2\%~\cite{SMP-12-008}, which can be directly translated into a
2.2\% uncertainty on the cross section normalization. It is fully
correlated across all \pt and mass bins.

\subsubsection{Unfolding uncertainty}

The unfolding correction is closely related to the dependence on the
\pt and $\Mjj$ resolution and the spectrum slope. For the
inclusive jet \pt spectrum it varies between 5\% and 10\%, while for
the dijet mass spectrum it ranges between 2\% and 5\%. The shape of
the unfolding correction and uncertainty as displayed in Fig.~\ref{fig_EXPunc} is
understood as follows: the resolution in the observable, \pt or
$\Mjj$, improves when going from low to high values. As a
consequence the effect of smearing is more pronounced in the lower \pt
or $\Mjj$ region. On the other hand the \pt and $\Mjj$
spectra become steeper when approaching the kinematic limit at high
\pt or $\Mjj$, leading again to a larger smearing
effect than observed at medium values.

The uncertainty introduced by the unfolding is caused by the modeling
of the jet \pt (dijet mass) resolution and the jet \pt (dijet mass)
spectrum in the simulation. In order to estimate the sensitivity of
the correction to these inputs, the jet \pt resolution is varied by
$\pm$10\% and the jet (dijet mass) spectrum slope by $\pm$5\%. The
former is motivated by the observed difference between data and
simulation in the jet energy resolution~\cite{JES}, and the latter is a
conservative estimate based on comparisons of the theoretical and
measured spectrum shapes. An additional constant 2\% uncertainty is
assigned to the dependence on the unfolding method. Overall, the
unfolding uncertainty is of the order of 3--4\%, and is fully correlated
across the \pt and mass bins.

\subsubsection{Other uncertainty sources}

The contributions from small trigger and jet identification
inefficiencies, time dependence of the jet \pt resolution, and
uncertainty on the trigger prescale factor have been shown to be much
smaller than 1\%. To account for these residual effects a conservative
uncertainty of 1\% is assigned to each jet \pt and dijet mass bin,
uncorrelated across the bins.
\section{Theoretical predictions}

The theoretical predictions for the jet cross sections consist of a
next-to-leading-order (NLO) QCD calculation and a nonperturbative (NP)
correction to account for the multiparton interactions (MPI) and
hadronization effects.

\subsection{NLO calculations}

The NLO calculations are performed using the NLOJet++ program
(v2.0.1)~\cite{NLO} within the framework of the fastNLO package
(v1.4)~\cite{fastNLO}. The renormalization and factorization scales
($\mu_R$ and $\mu_F$) for the inclusive and dijet measurements are
identified with the jet \pt and the average transverse momentum
$\pt^\text{ave}$ of the two jets, respectively. The NLO calculation is
performed using five different PDF sets: CT10~\cite{CT10},
MSTW2008NLO~\cite{MSTW}, NNPDF2.1~\cite{NNPDF},
HERAPDF1.5~\cite{HERA}, and ABKM09~\cite{ABKM} at the corresponding
default values of the strong coupling constant $\alpS(M_Z)= 0.1180,
0.120$, $0.119$, $0.1176$, and $0.1179$, respectively.

\subsection{Systematic uncertainties}

\begin{figure*}[htbp]
  \centering
  \includegraphics[width=0.40\textwidth]{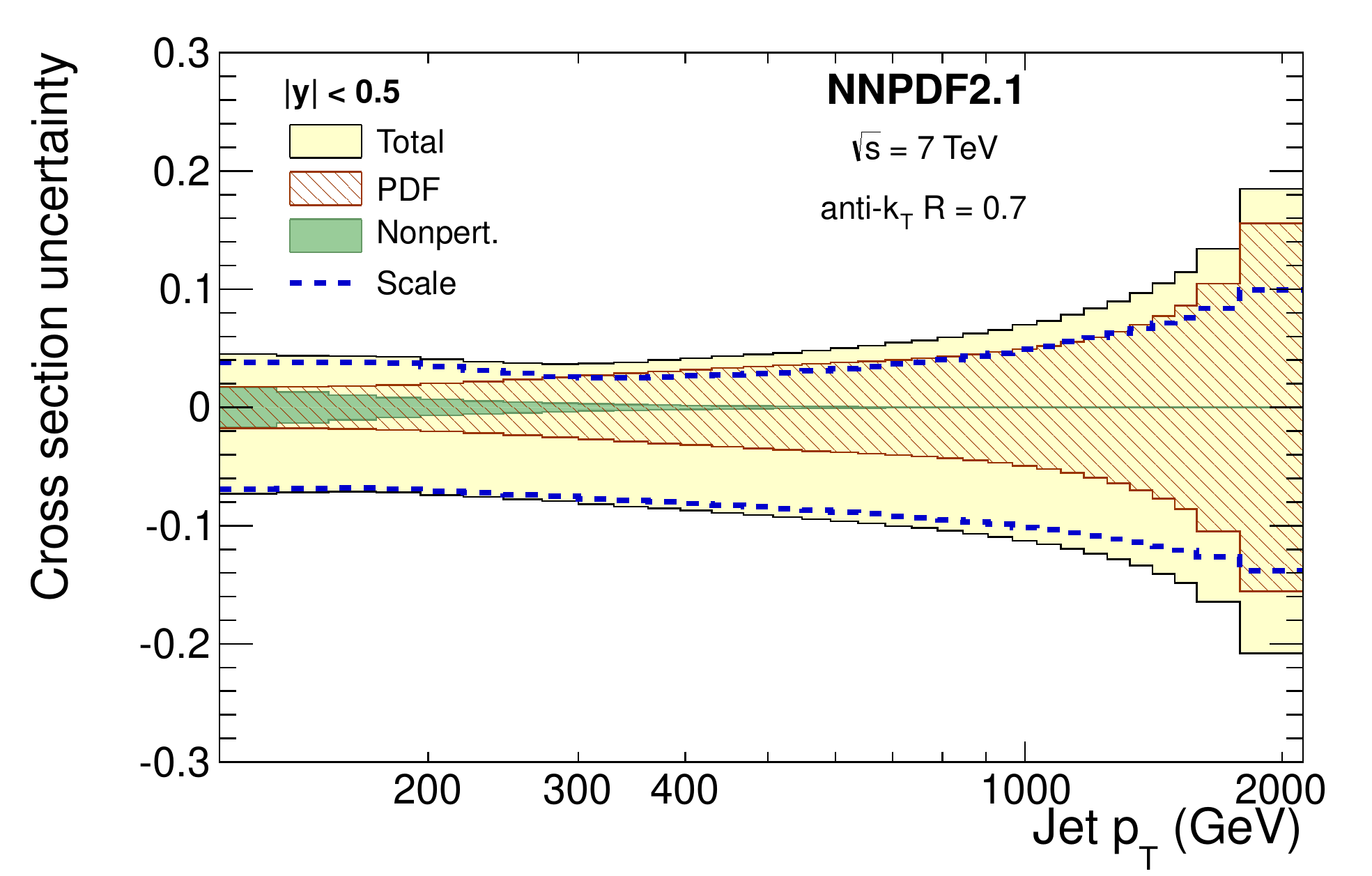}
  \includegraphics[width=0.40\textwidth]{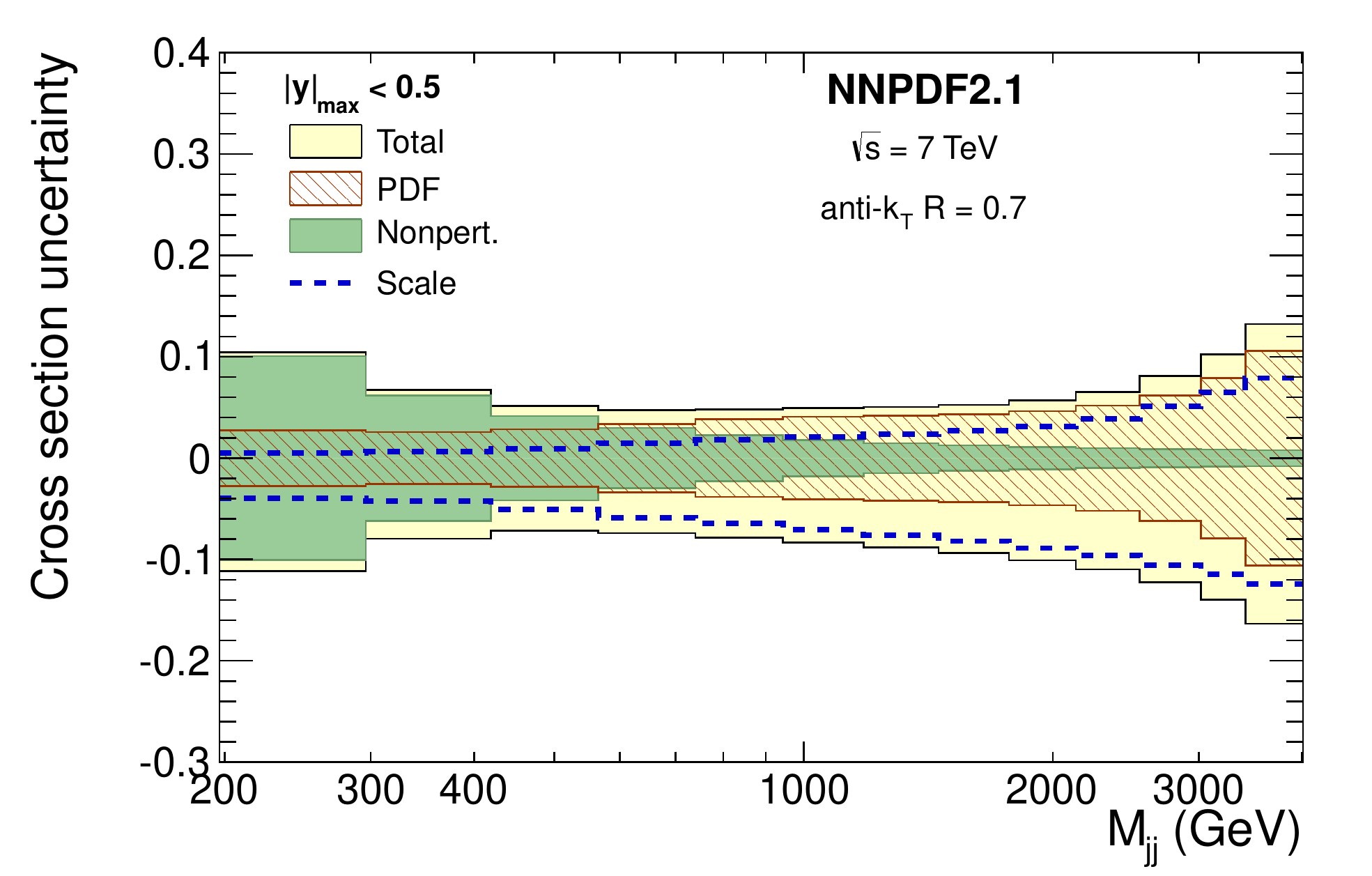}
  \includegraphics[width=0.40\textwidth]{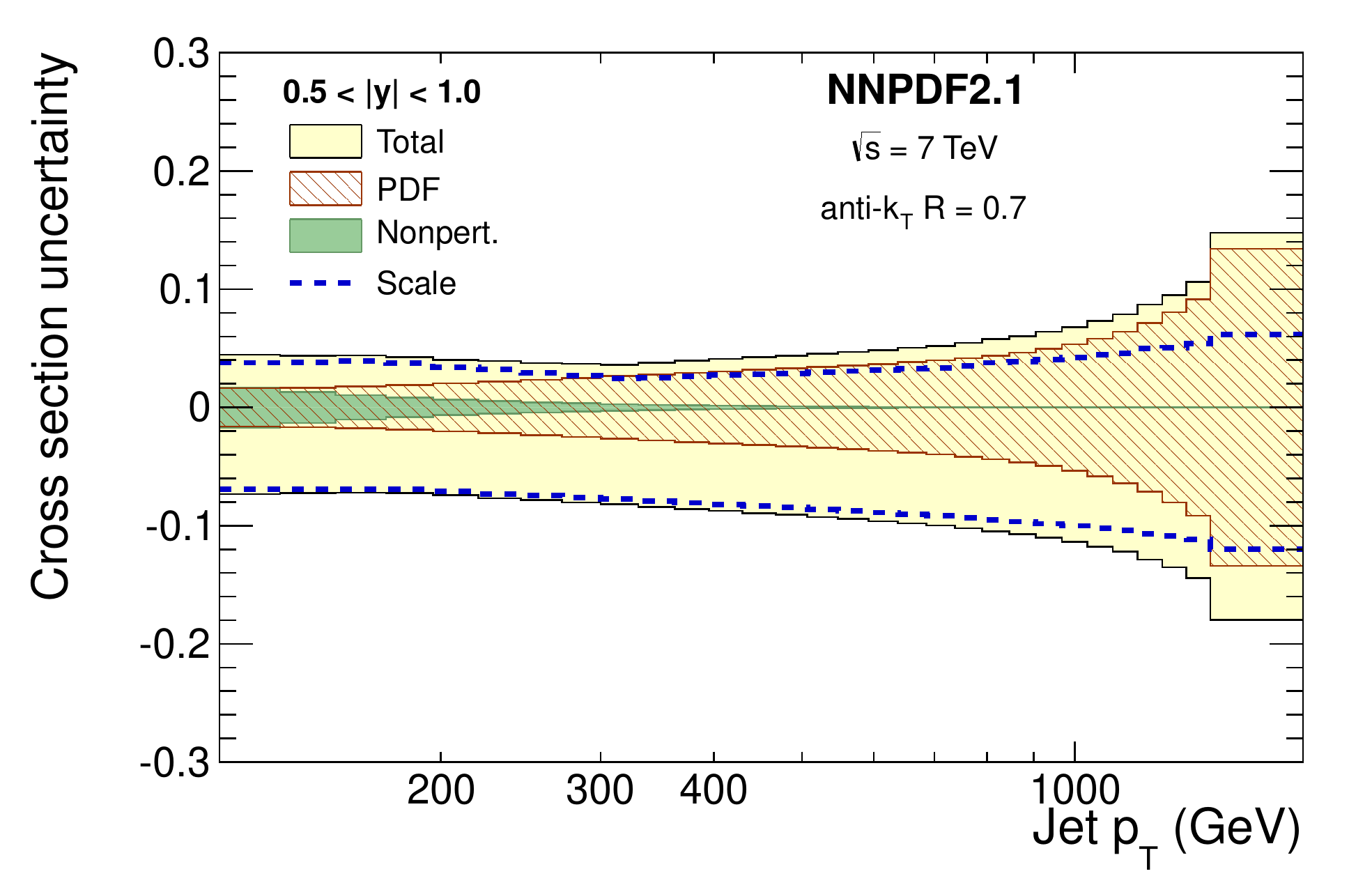}
  \includegraphics[width=0.40\textwidth]{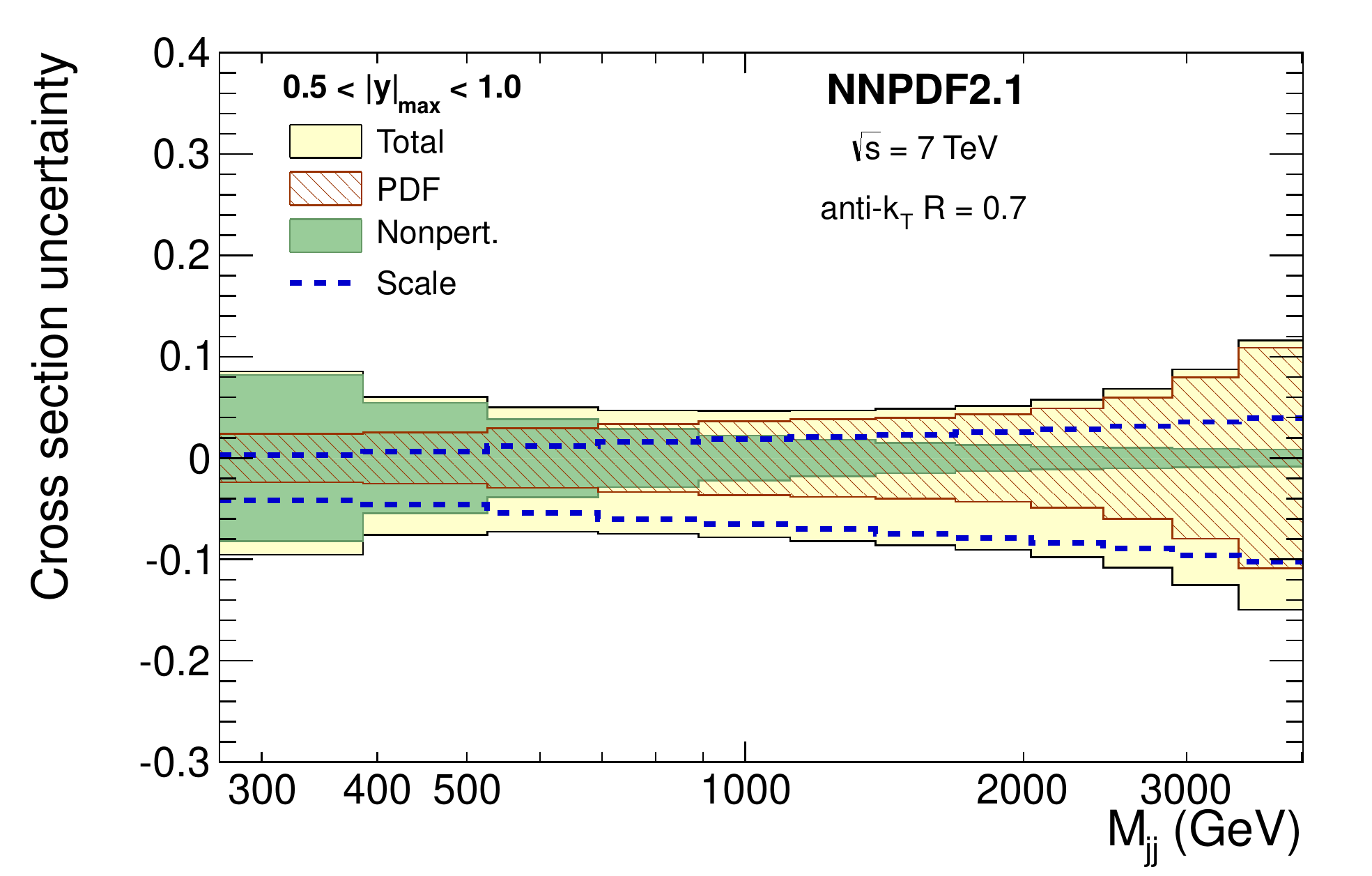}
  \includegraphics[width=0.40\textwidth]{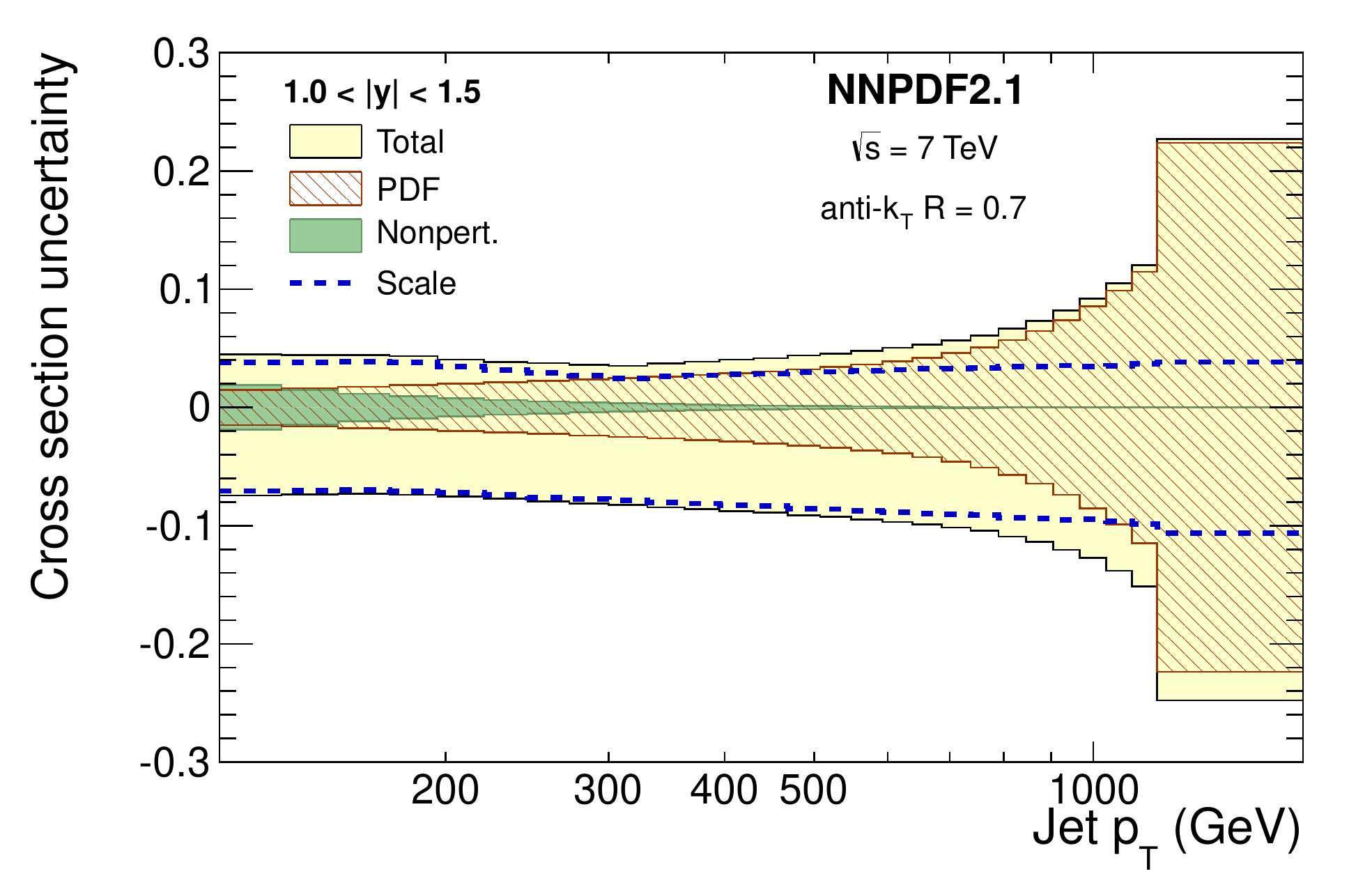}
  \includegraphics[width=0.40\textwidth]{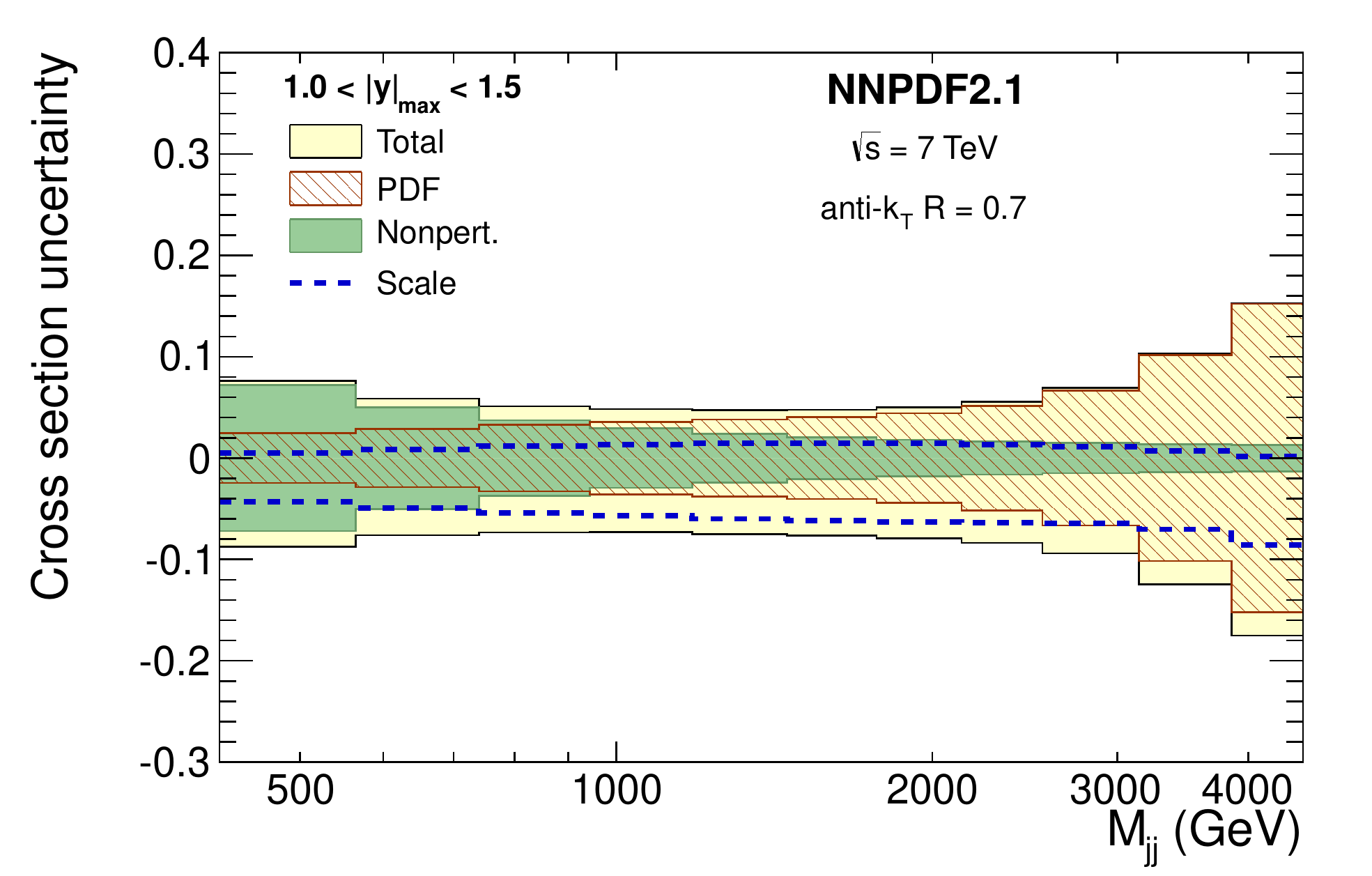}
  \includegraphics[width=0.40\textwidth]{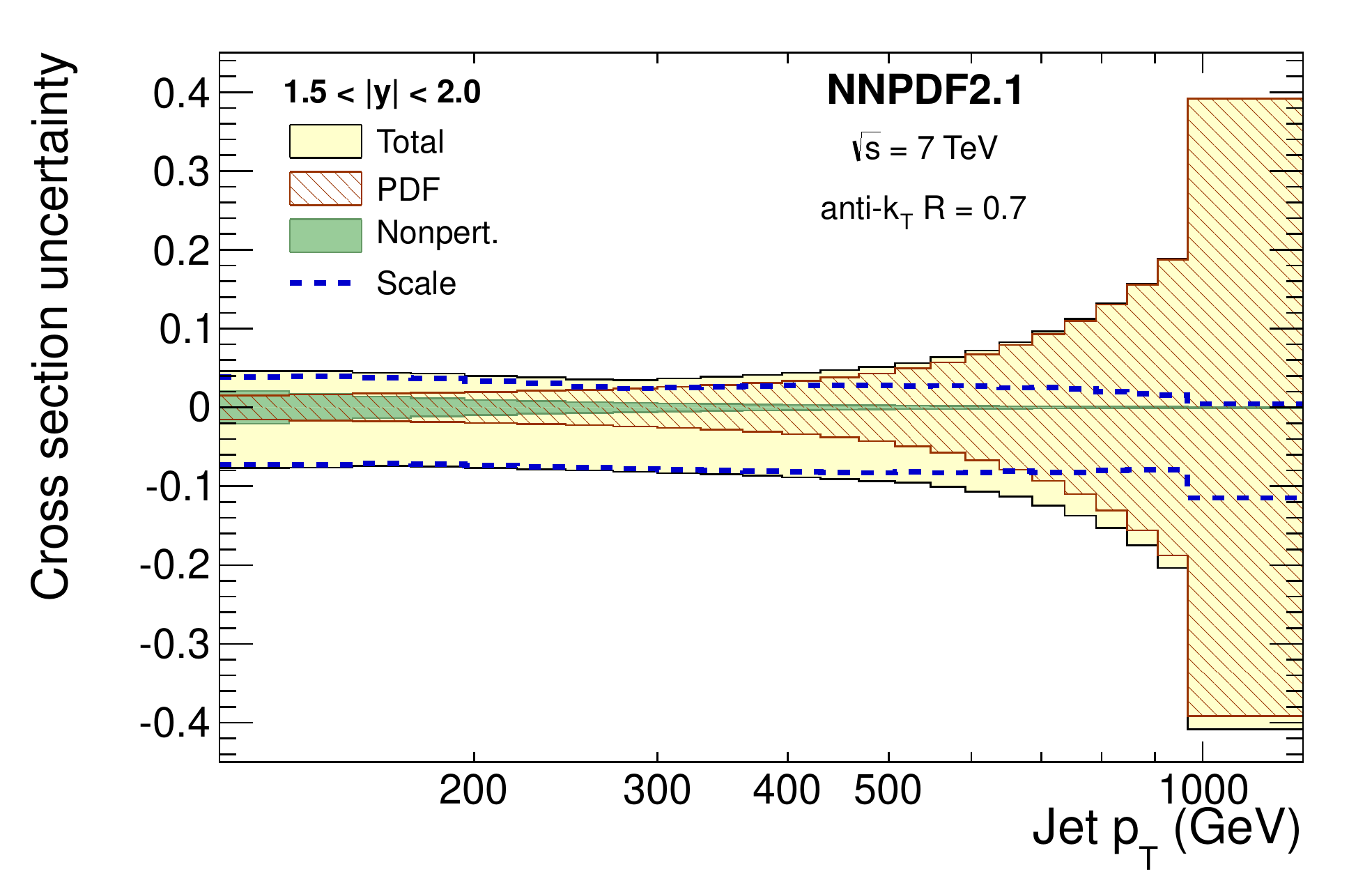}
  \includegraphics[width=0.40\textwidth]{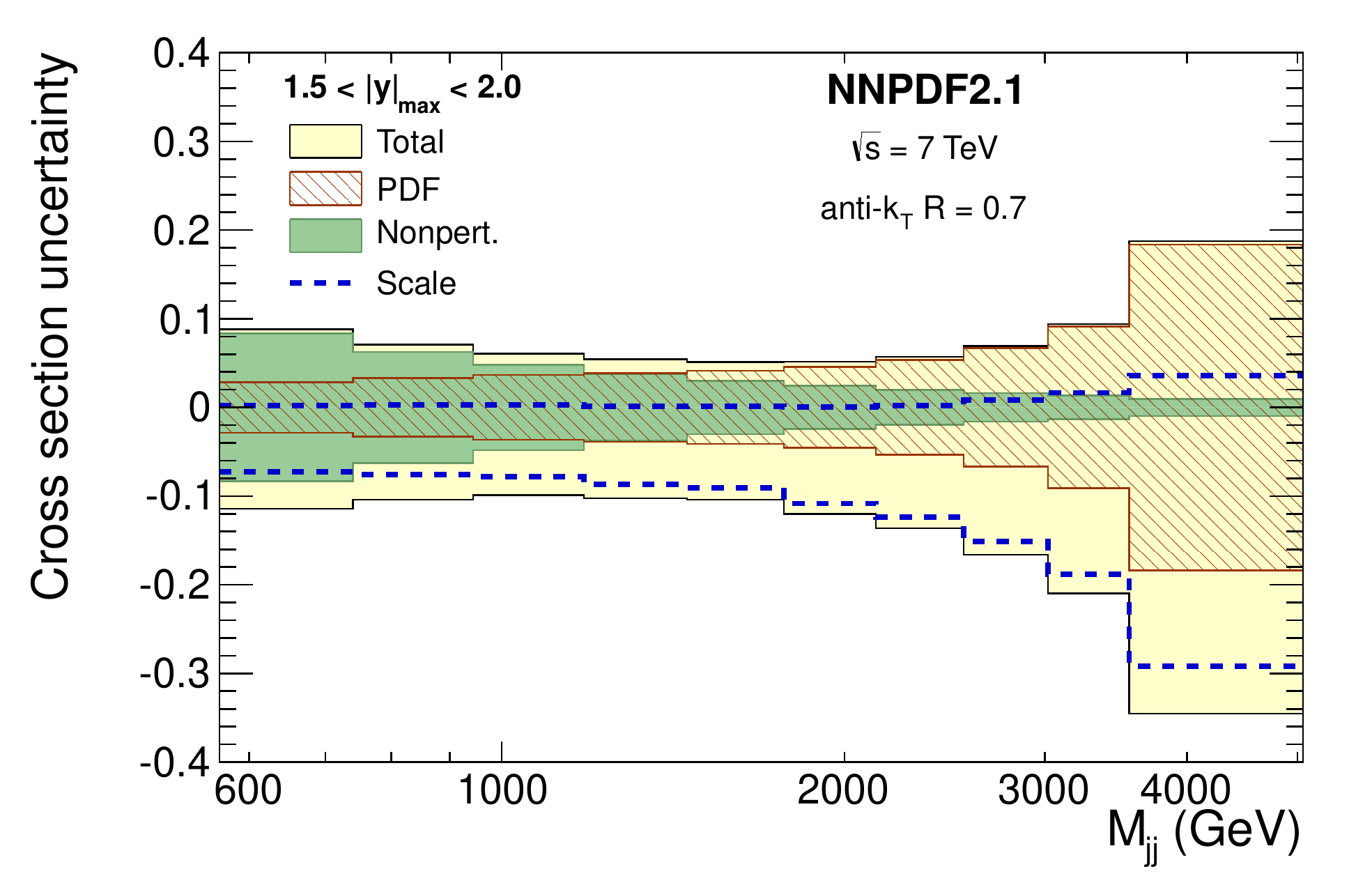}
  \includegraphics[width=0.40\textwidth]{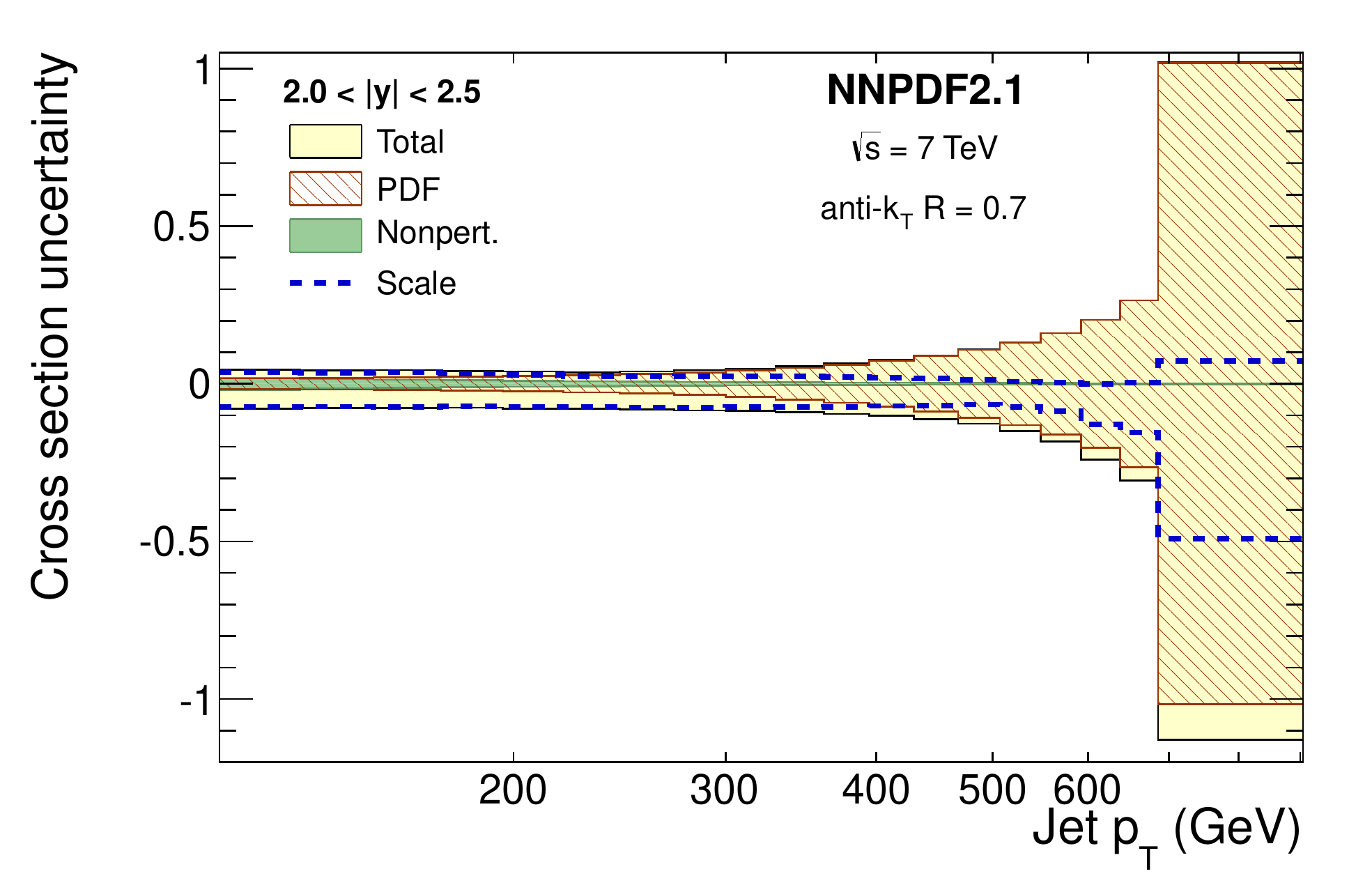}
  \includegraphics[width=0.40\textwidth]{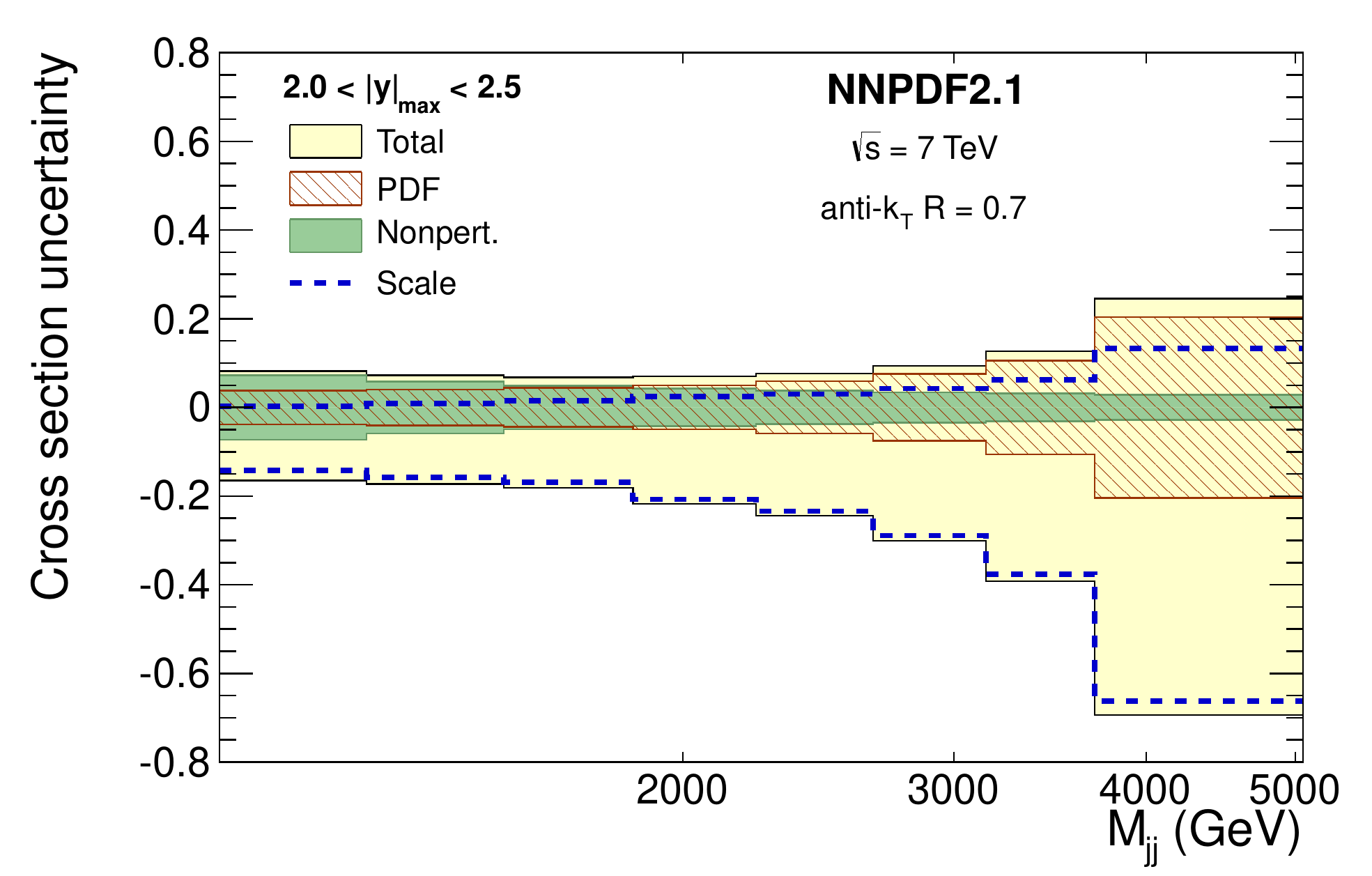}
  \caption{Effect of the relative theoretical uncertainties for the
    inclusive jet (left column) and dijet (right column) cross section
    measurements for all five $|y|$ and $|y|_\text{max}$
    bins, respectively. The upward and downward uncertainties are estimated separately.}
  \label{fig_theUnc}
\end{figure*}

The PDF variation introduces uncertainties in the theoretical
prediction of up to 30\%, while the variation of $\alpS(M_Z)$ by
${\pm}0.001$ introduces an additional 1--2\% uncertainty.  The
uncertainty due to the choice of factorization and renormalization
scales is estimated as the maximum deviation at the six points
($\mu_F/\mu,\mu_R/\mu) = (0.5,0.5)$, $(2,2)$, $(1,0.5)$, $(1,2)$,
$(0.5,1)$, $(2,1)$, where $\mu=\pt$ (inclusive) or
$\mu=\pt^\text{ave}$ (dijet). An additional uncertainty of at most
$10\%$ is caused by the nonperturbative correction. The scale
uncertainty ranges from 5\% to 10\% for $|y|<1.5$ but increases to
40\% for the outer $|y|$ bins and for high dijet masses and jet \pt.
Overall, the PDF uncertainty is dominant for the high \pt and high dijet
mass regions.  Figure~\ref{fig_theUnc} shows the effect of the
systematic uncertainties for the two observables in all rapidity bins
and for the NNPDF2.1 PDF set.

\subsection{Nonperturbative corrections}

The nonperturbative effects are estimated from the simulation, using
the event generators \PYTHIA{6} (tune Z2) and {\HERWIG{++} 2.4.2}~\cite{HERWIG}. (The \PYTHIA{6} Z2 tune is identical to the Z1 tune described
  in~\cite{D6T} except that Z2 uses the CTEQ6L PDF while Z1 uses
  CTEQ5L.) These models are
representative of the possible values of the nonperturbative
corrections, due to their different physics descriptions. The
nonperturbative correction is defined as the ratio of the cross
section predicted with the nominal generator settings divided by the
cross section predicted with the MPI and hadronization switched off.
The central value of the nonperturbative correction is calculated from
the average of the two models considered, and ranges from 1\% to
20\%, being larger in the dijet spectrum because of the involvement of
lower \pt jets.
\section{Results}

\begin{figure}[hbtp]
  \begin{center}
    \includegraphics[width=0.49\textwidth]{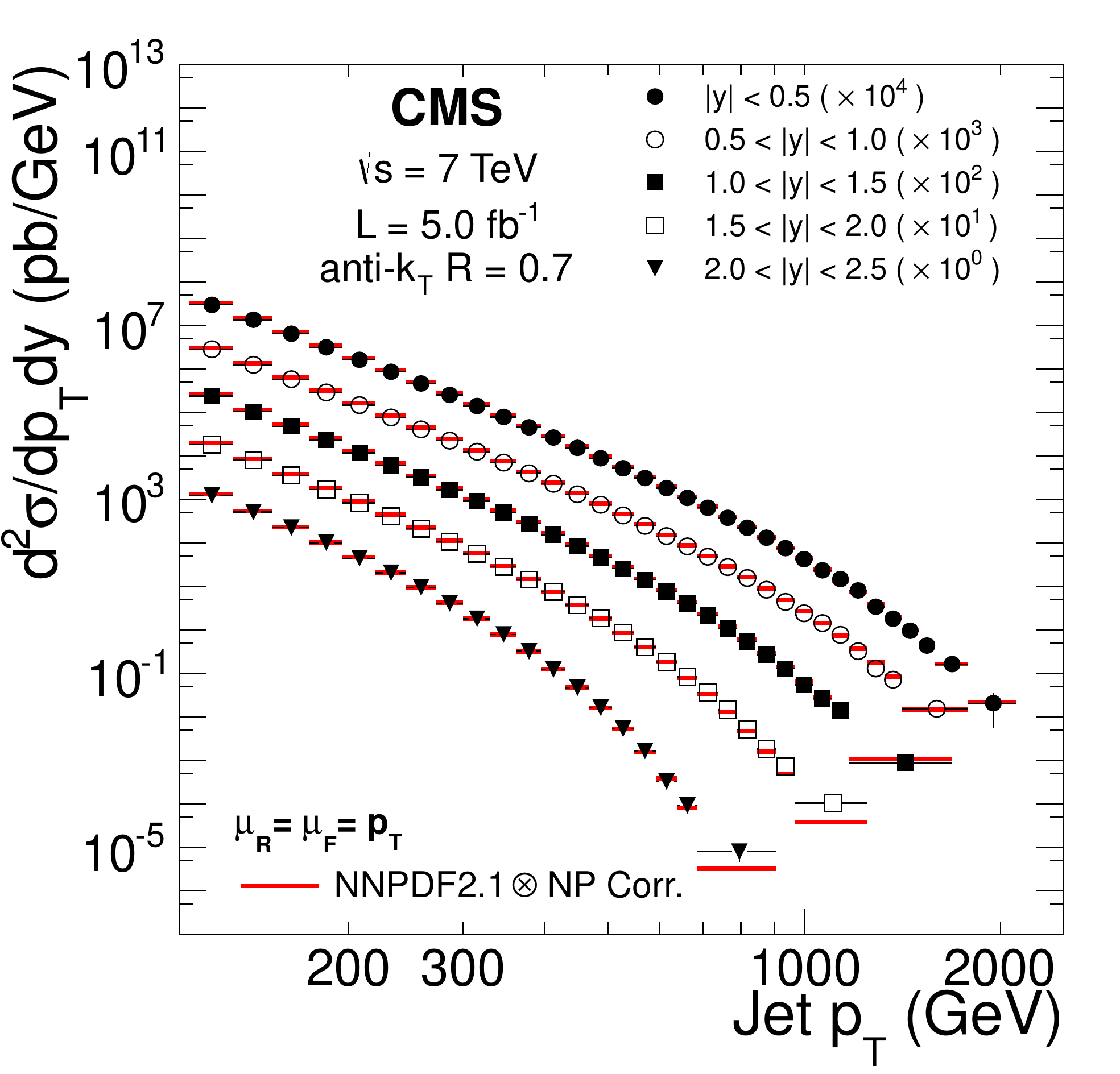}
    \includegraphics[width=0.49\textwidth]{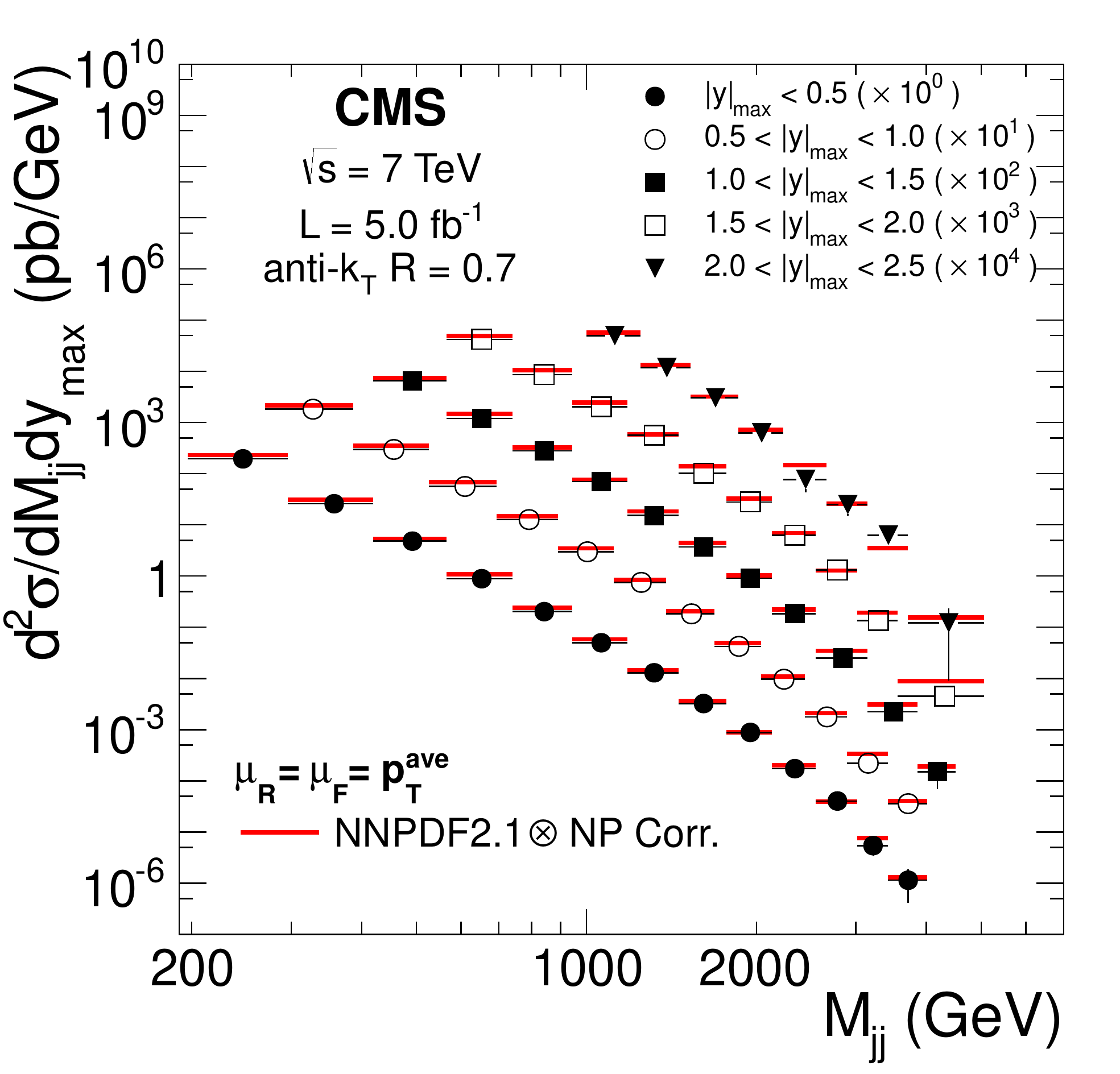}
    \caption{Inclusive jet (\cmsLeft) and dijet (\cmsRight) cross sections for
      the five different rapidity bins, for data (markers) and theory
      (thick lines) using the NNPDF2.1 PDF set.}
    \label{fig:spectra}
  \end{center}
\end{figure}

The unfolded inclusive jet and dijet spectra are shown in
Fig.~\ref{fig:spectra}, compared to the theoretical predictions.

\begin{figure*}[hbtp]
  \begin{center}
    \includegraphics[width=0.48\textwidth]{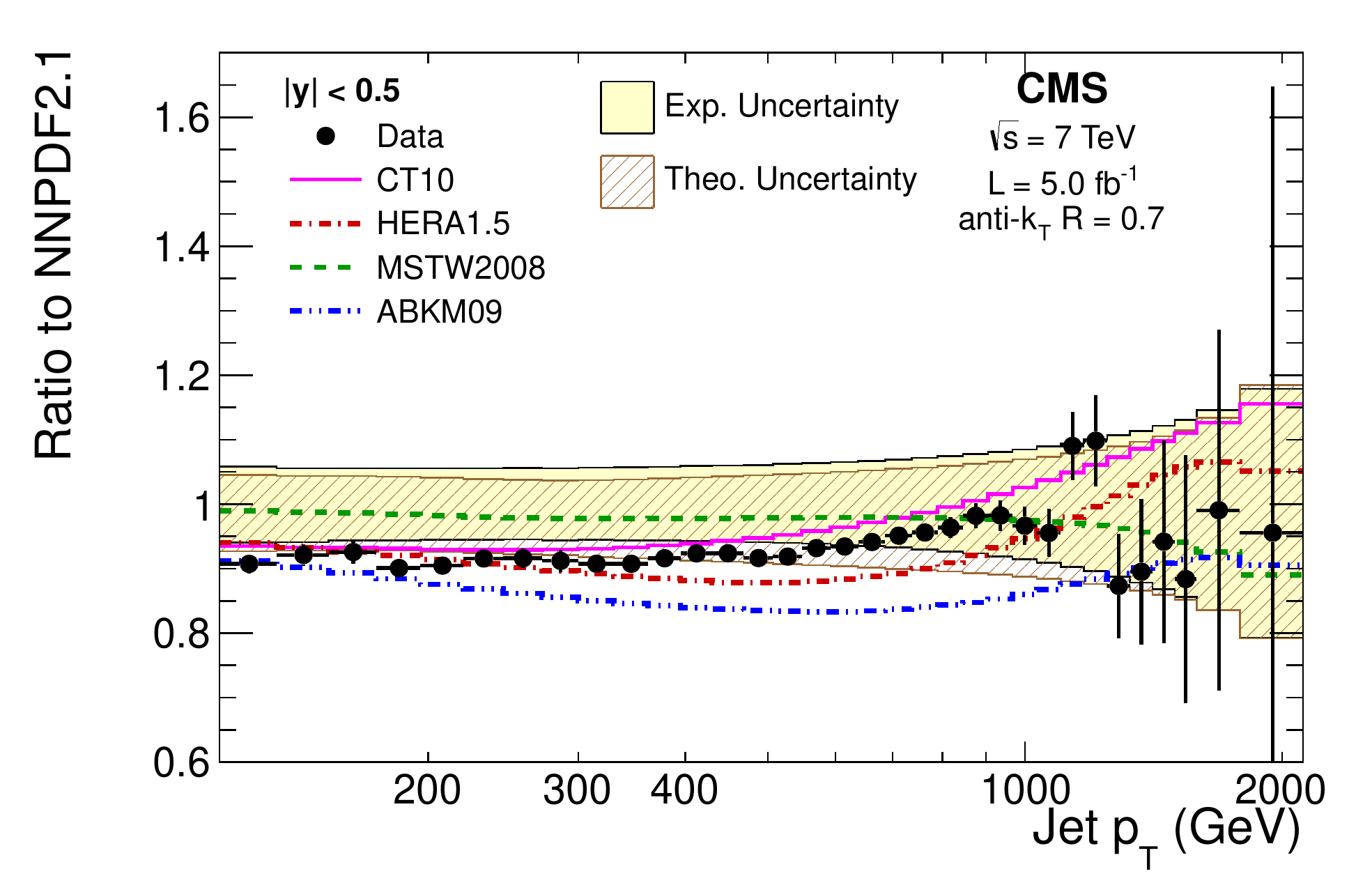}
    \includegraphics[width=0.48\textwidth]{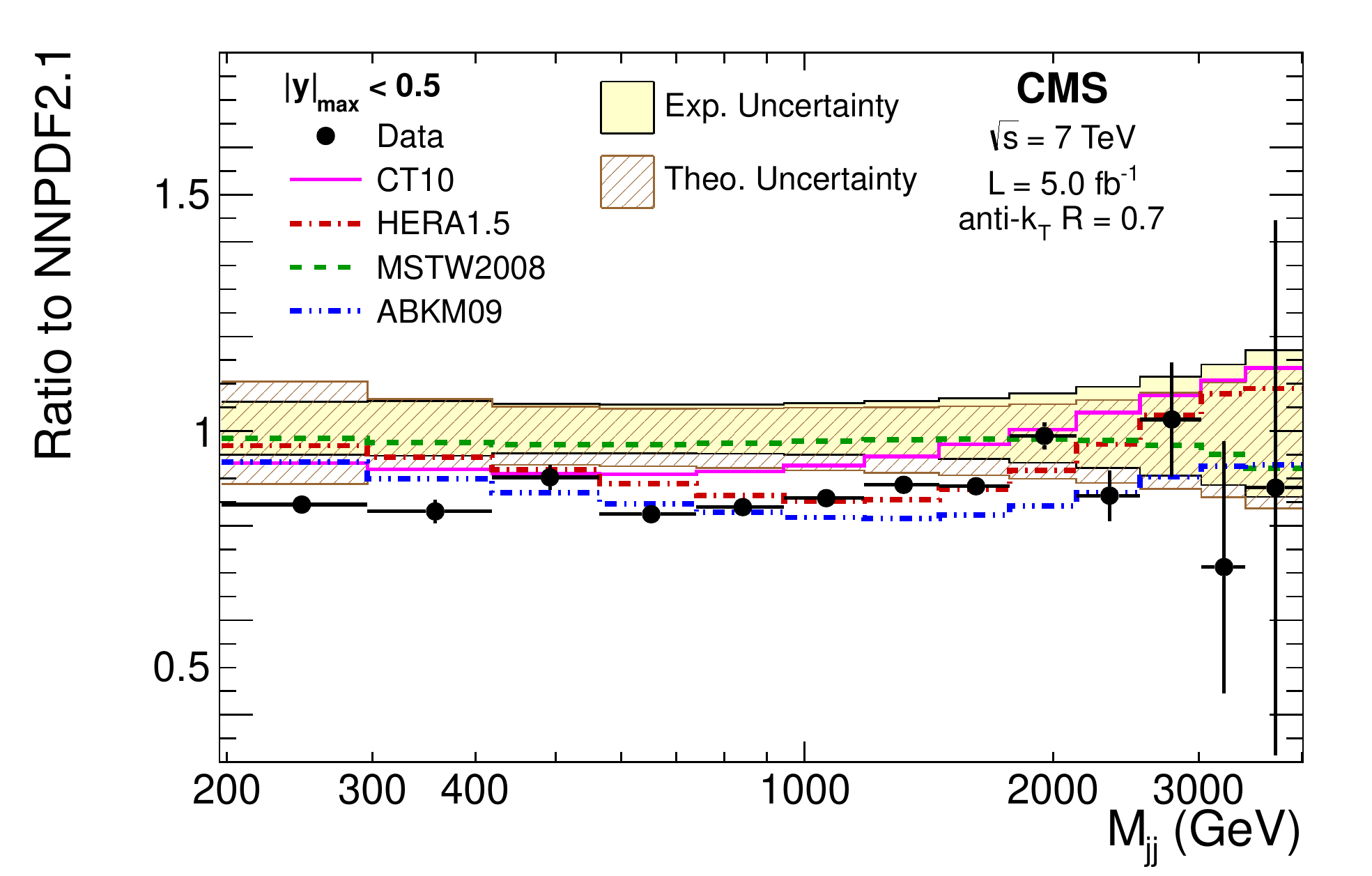}
    \includegraphics[width=0.48\textwidth]{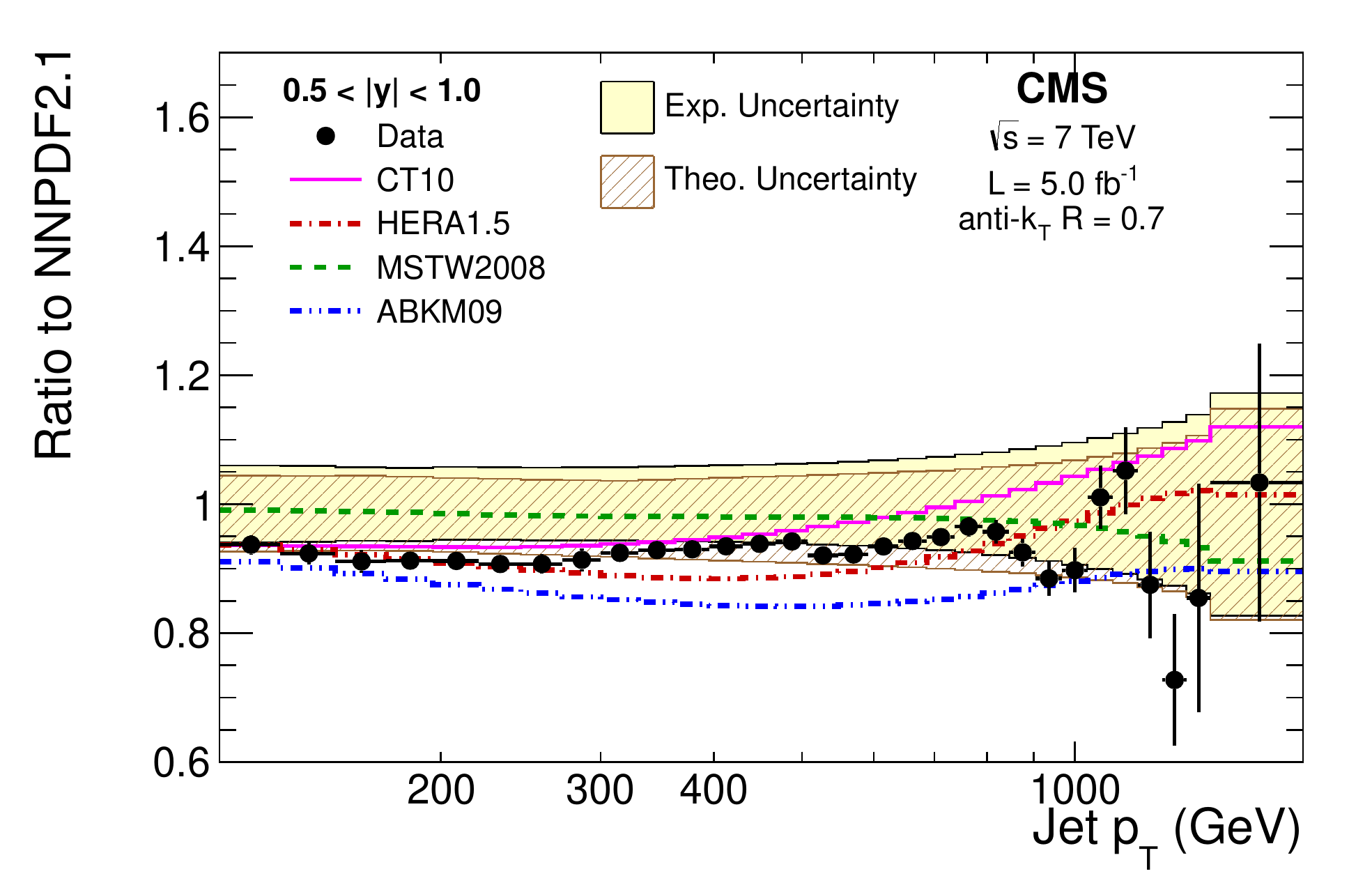}
    \includegraphics[width=0.48\textwidth]{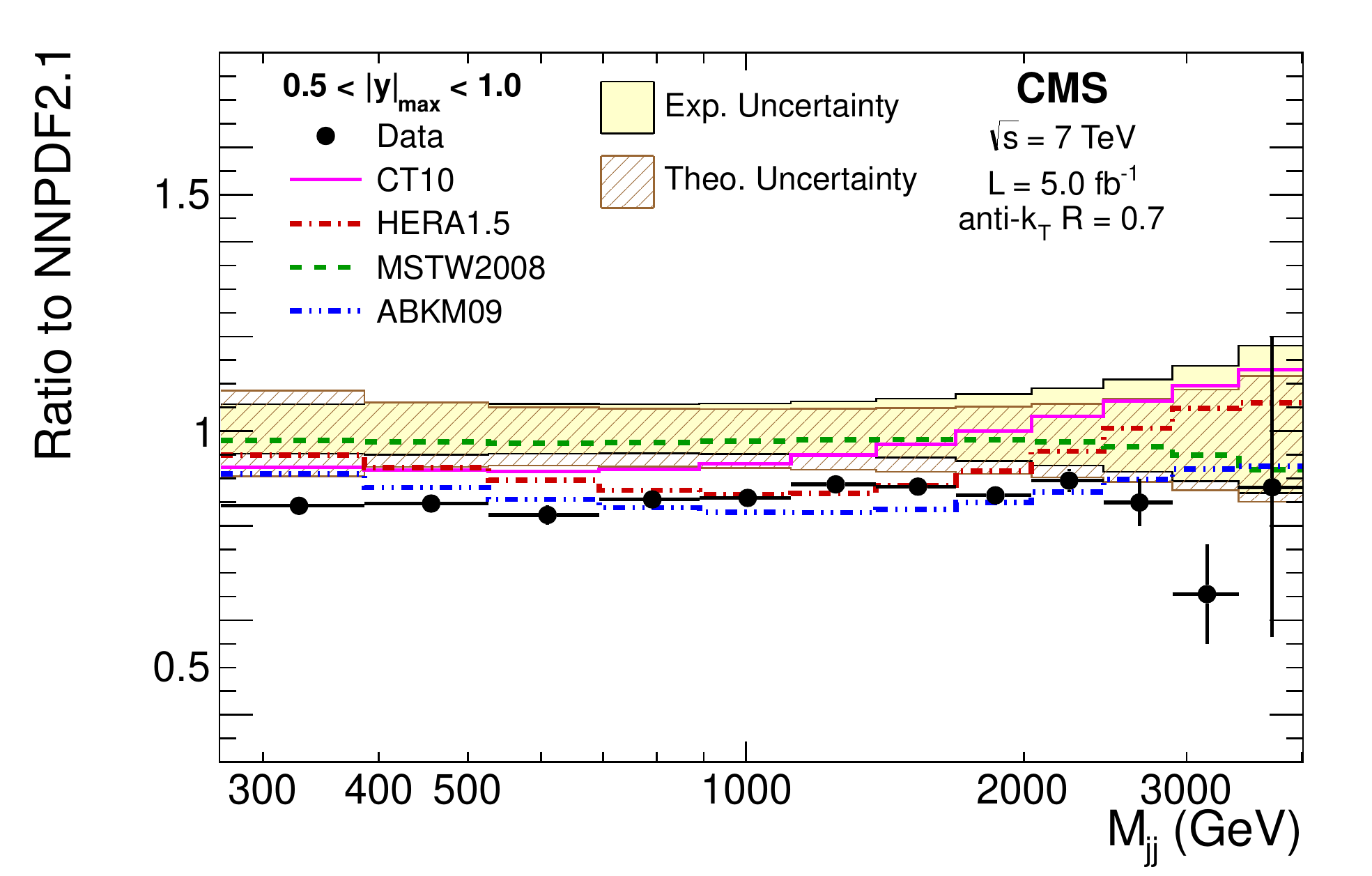}
    \includegraphics[width=0.48\textwidth]{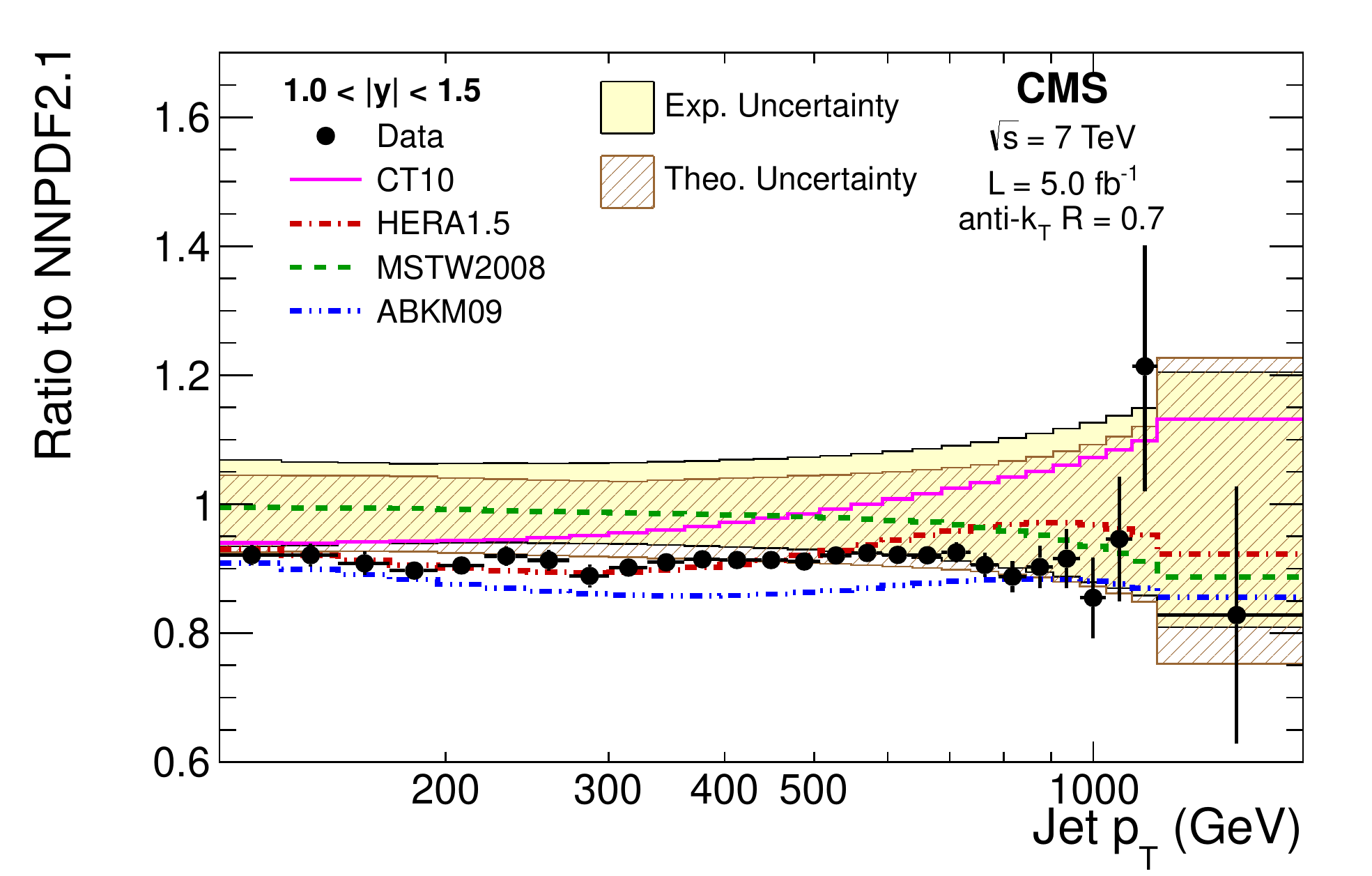}
    \includegraphics[width=0.48\textwidth]{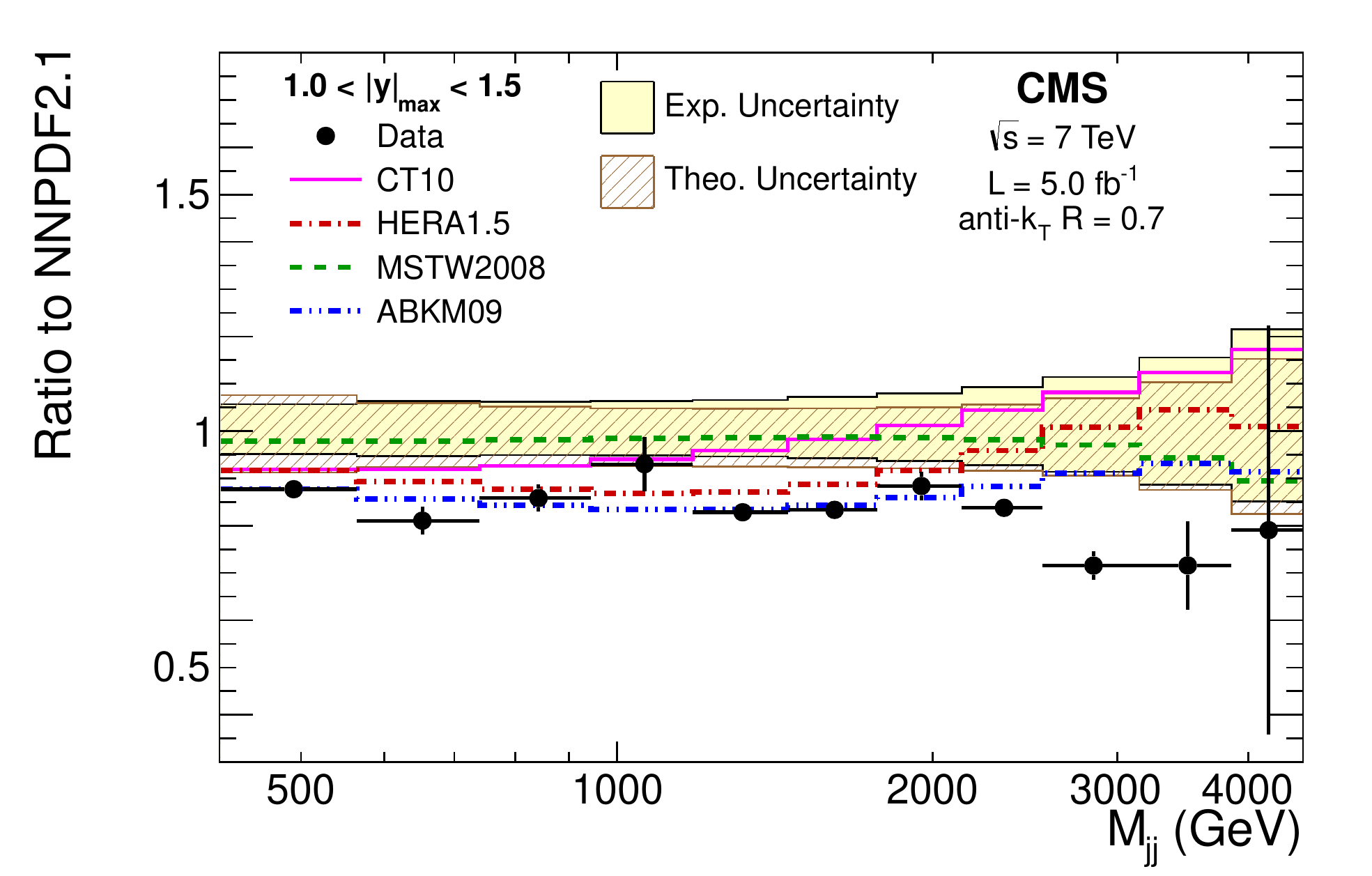}
    \caption{Ratio of inclusive jet (left) and dijet (right) cross
      sections to the theoretical prediction using the central value of the
      NNPDF2.1 PDF set for the first three $|y|$ and $|y|_\text{max}$ bins respectively.  The
      solid histograms show the ratio of the
      cross sections calculated with the other PDF sets to that
      calculated with NNPDF2.1.  The experimental and theoretical
      systematic uncertainties are represented by the continuous and
      hatched bands, respectively.}
    \label{fig:RatioALL}
  \end{center}
\end{figure*}

\begin{figure*}[hbtp]
  \begin{center}
    \includegraphics[width=0.48\textwidth]{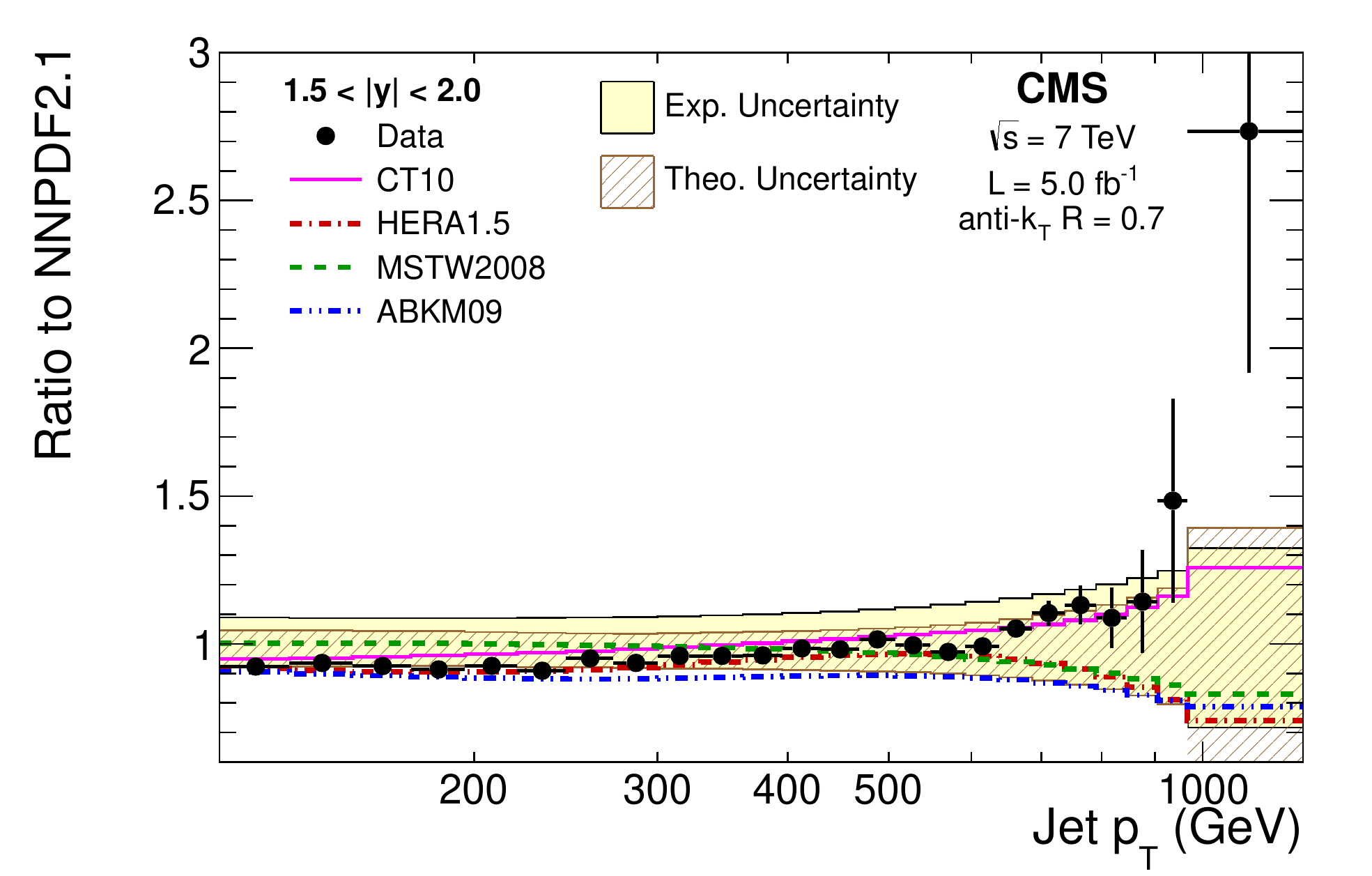}
    \includegraphics[width=0.48\textwidth]{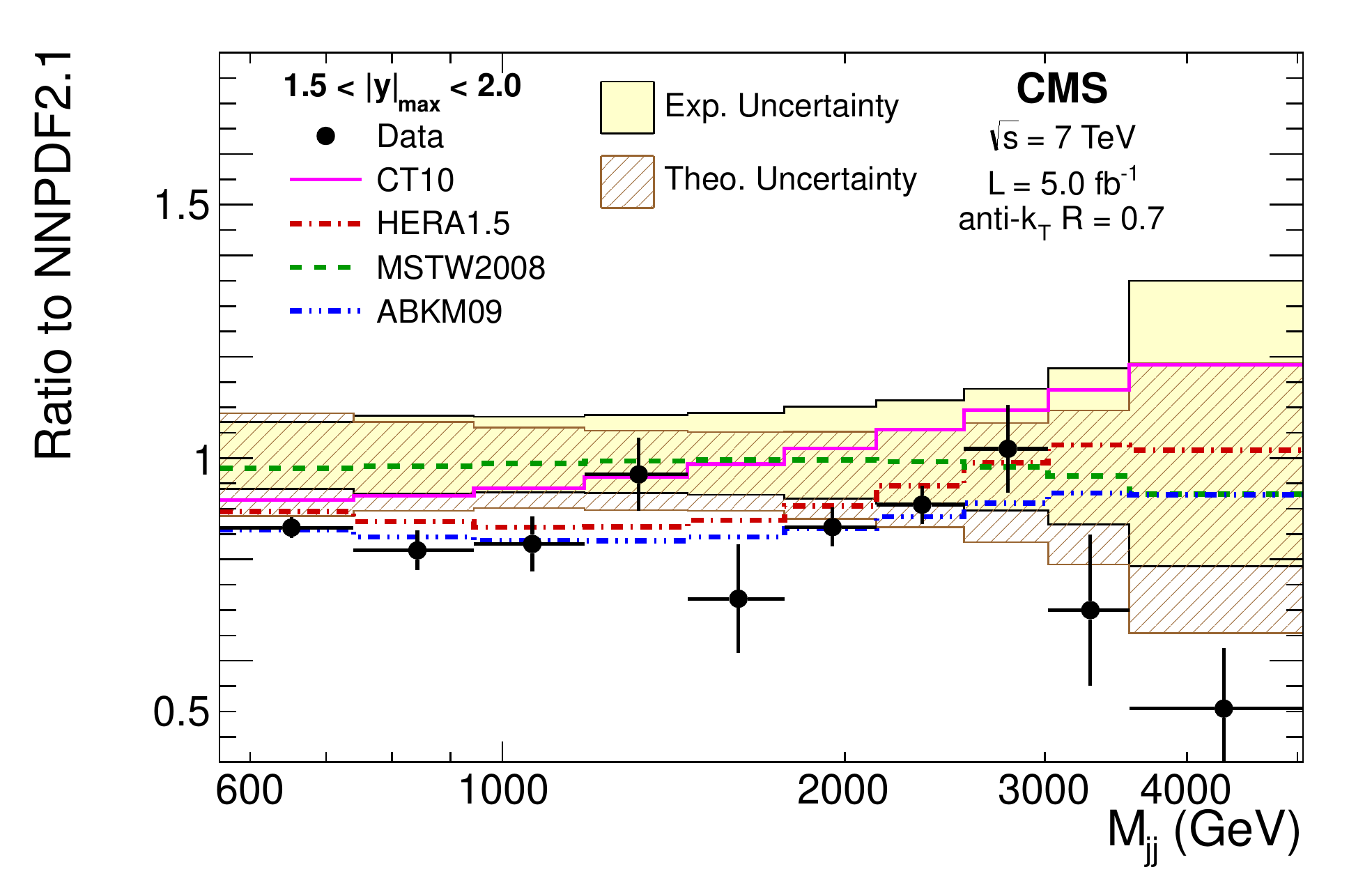}
    \includegraphics[width=0.48\textwidth]{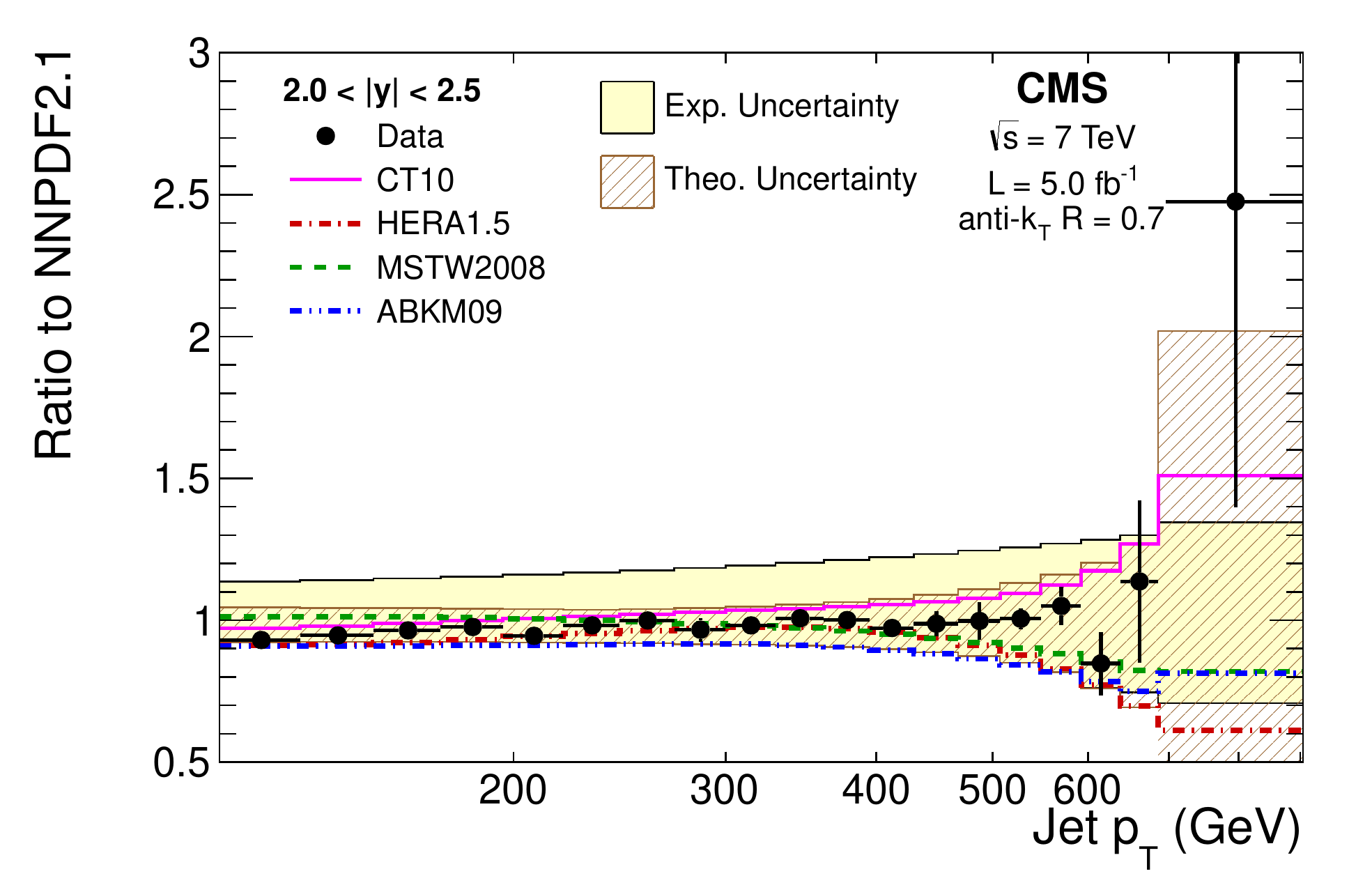}
    \includegraphics[width=0.48\textwidth]{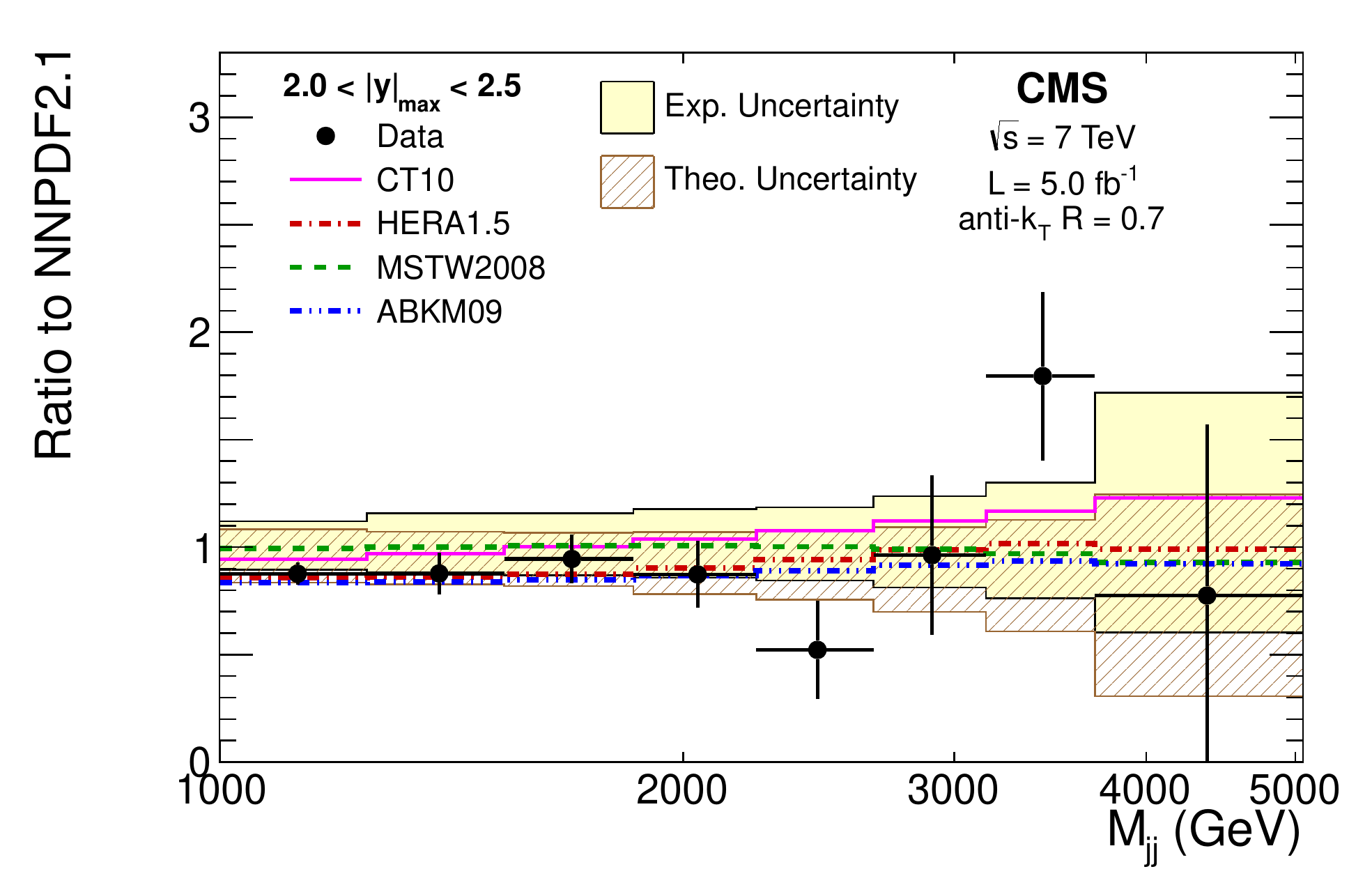}
    \caption{Ratio of inclusive jet (left) and dijet (right) cross
      sections to the theoretical prediction using the central value of the
      NNPDF2.1 PDF set for the last two $|y|$ and $|y|_\text{max}$ bins respectively.  The
       solid histograms show the ratio of the
      cross sections calculated with the other PDF sets to that
      calculated with NNPDF2.1.  The experimental and theoretical
      systematic uncertainties are represented by the continuous and
      hatched bands, respectively.}
    \label{fig:RatioALL1}
  \end{center}
\end{figure*}

To compare the CMS data and the theoretical prediction, the ratio of
the two is taken. Figures~\ref{fig:RatioALL}--\ref{fig:RatioALL1} show
this ratio using the central value of the NNPDF2.1 PDF set,
accompanied by the total experimental and theoretical uncertainties.
The theoretical uncertainties vary considerably among the different
PDF sets, and in particular in the high-\pt and high-$\Mjj$
region. The experimental uncertainty is comparable to the theoretical
uncertainty. The additional curves represent the ratio of the central values of the other
PDF sets to NNPDF2.1.  Agreement is observed between
data and theory in all rapidity bins, given the statistical and
systematic uncertainties, with the various theoretical predictions
showing differences of the order of 10\%.

\section{Summary}

Measurements of the double-differential inclusive jet and dijet cross
sections are presented using 5.0\fbinv of data collected with
the CMS detector in proton-proton collisions at $\sqrt{s}=7\TeV$.  The
measurements cover the jet \pt range from 0.1\TeV to 2\TeV, and
the dijet-mass range from 0.3\TeV to 5\TeV in five rapidity bins
up to $|y|=2.5$. The measured cross sections agree with the
predictions of perturbative QCD at next-to-leading order obtained with
five different PDF sets.  Theoretical and experimental uncertainties
are comparable, even at the limits of the experimental phase space,
so these results may be used to constrain global PDF fits.

\section*{Acknowledgements}
\hyphenation{Bundes-ministerium Forschungs-gemeinschaft Forschungs-zentren} We congratulate
our colleagues in the CERN accelerator departments for the excellent performance of the LHC
and thank the technical and administrative staffs at CERN and at other CMS institutes for
their contributions to the success of the CMS effort. In addition, we gratefully acknowledge
the computing centers and personnel of the Worldwide LHC Computing Grid for delivering so
effectively the computing infrastructure essential to our analyses. Finally, we acknowledge
the enduring support for the construction and operation of the LHC and the CMS detector
provided by the following funding agencies: the Austrian Federal Ministry of Science and
Research; the Belgian Fonds de la Recherche Scientifique, and Fonds voor Wetenschappelijk
Onderzoek; the Brazilian Funding Agencies (CNPq, CAPES, FAPERJ, and FAPESP); the Bulgarian
Ministry of Education, Youth and Science; CERN; the Chinese Academy of Sciences, Ministry of
Science and Technology, and National Natural Science Foundation of China; the Colombian
Funding Agency (COLCIENCIAS); the Croatian Ministry of Science, Education and Sport; the
Research Promotion Foundation, Cyprus; the Ministry of Education and Research, Recurrent
financing contract SF0690030s09 and European Regional Development Fund, Estonia; the Academy
of Finland, Finnish Ministry of Education and Culture, and Helsinki Institute of Physics; the
Institut National de Physique Nucl\'eaire et de Physique des Particules~/~CNRS, and
Commissariat \`a l'\'Energie Atomique et aux \'Energies Alternatives~/~CEA, France; the
Bundesministerium f\"ur Bildung und Forschung, Deutsche Forschungsgemeinschaft, and
Helmholtz-Gemeinschaft Deutscher Forschungszentren, Germany; the General Secretariat for
Research and Technology, Greece; the National Scientific Research Foundation, and National
Office for Research and Technology, Hungary; the Department of Atomic Energy and the
Department of Science and Technology, India; the Institute for Studies in Theoretical Physics
and Mathematics, Iran; the Science Foundation, Ireland; the Istituto Nazionale di Fisica
Nucleare, Italy; the Korean Ministry of Education, Science and Technology and the World Class
University program of NRF, Republic of Korea; the Lithuanian Academy of Sciences; the Mexican
Funding Agencies (CINVESTAV, CONACYT, SEP, and UASLP-FAI); the Ministry of Science and
Innovation, New Zealand; the Pakistan Atomic Energy Commission; the Ministry of Science and
Higher Education and the National Science Centre, Poland; the Funda\c{c}\~ao para a Ci\^encia
e a Tecnologia, Portugal; JINR (Armenia, Belarus, Georgia, Ukraine, Uzbekistan); the Ministry
of Education and Science of the Russian Federation, the Federal Agency of Atomic Energy of the
Russian Federation, Russian Academy of Sciences, and the Russian Foundation for Basic
Research; the Ministry of Science and Technological Development of Serbia; the Secretar\'{\i}a
de Estado de Investigaci\'on, Desarrollo e Innovaci\'on and Programa Consolider-Ingenio 2010,
Spain; the Swiss Funding Agencies (ETH Board, ETH Zurich, PSI, SNF, UniZH, Canton Zurich, and
SER); the National Science Council, Taipei; the Thailand Center of Excellence in Physics, the
Institute for the Promotion of Teaching Science and Technology of Thailand and the National
Science and Technology Development Agency of Thailand; the Scientific and Technical Research
Council of Turkey, and Turkish Atomic Energy Authority; the Science and Technology Facilities
Council, UK; the US Department of Energy, and the US National Science Foundation.
Individuals have received support from the Marie-Curie programme and the European Research
Council (European Union); the Leventis Foundation; the A. P. Sloan Foundation; the Alexander
von Humboldt Foundation; the Belgian Federal Science Policy Office; the Fonds pour la
Formation \`a la Recherche dans l'Industrie et dans l'Agriculture (FRIA-Belgium); the
Agentschap voor Innovatie door Wetenschap en Technologie (IWT-Belgium); the Ministry of
Education, Youth and Sports (MEYS) of Czech Republic; the Council of Science and Industrial
Research, India; the Compagnia di San Paolo (Torino); and the HOMING PLUS programme of
Foundation for Polish Science, cofinanced from European Union, Regional Development Fund.

\bibliography{auto_generated}   
\cleardoublepage \appendix\section{The CMS Collaboration \label{app:collab}}\begin{sloppypar}\hyphenpenalty=5000\widowpenalty=500\clubpenalty=5000\textbf{Yerevan Physics Institute,  Yerevan,  Armenia}\\*[0pt]
S.~Chatrchyan, V.~Khachatryan, A.M.~Sirunyan, A.~Tumasyan
\vskip\cmsinstskip
\textbf{Institut f\"{u}r Hochenergiephysik der OeAW,  Wien,  Austria}\\*[0pt]
W.~Adam, E.~Aguilo, T.~Bergauer, M.~Dragicevic, J.~Er\"{o}, C.~Fabjan\cmsAuthorMark{1}, M.~Friedl, R.~Fr\"{u}hwirth\cmsAuthorMark{1}, V.M.~Ghete, J.~Hammer, N.~H\"{o}rmann, J.~Hrubec, M.~Jeitler\cmsAuthorMark{1}, W.~Kiesenhofer, V.~Kn\"{u}nz, M.~Krammer\cmsAuthorMark{1}, I.~Kr\"{a}tschmer, D.~Liko, I.~Mikulec, M.~Pernicka$^{\textrm{\dag}}$, B.~Rahbaran, C.~Rohringer, H.~Rohringer, R.~Sch\"{o}fbeck, J.~Strauss, A.~Taurok, W.~Waltenberger, C.-E.~Wulz\cmsAuthorMark{1}
% ...
\vskip\cmsinstskip
\dag:~Deceased\\
1:~~Also at Vienna University of Technology, Vienna, Austria\\
% ...

\end{sloppypar}
\end{document}